# EVOLUTIONARY ANALYSIS OF BIOLOGICAL EXCITABILITY

AR. ASHOK PALLANIAPPAN[1,2*]

[1]Madras Institute of Systems Biology

Chennai 600 040

[2]Laboratory of Translational Research

Chettinad Medical University

Chennai 608 002

India

ERIC G. JAKOBSSON

Professor Emeritus

University of Illinois at Urbana-Champaign

Urbana, IL 61801

U.S.A.

[*]Corresponding author
Phone: +91.44.42845265
http:\\www.misb.in

**ABSTRACT**

Excitability is an attribute of life, and is a driving force in the descent of complexity. Cellular electrical activity as realized by membrane proteins that act as either channels or transporters is the basis of excitability. Electrical signaling is mediated by a wave of action potentials, which consist of synchronous redistribution of ionic gradients down ion channels. Ion channels select for the passage of a particular ion species. Potassium ion channels are gated by a variety of stimuli, including membrane voltage. Sodium and calcium channels are gated only by membrane voltage, suggesting the conservative argument that voltage-gated potassium channels are the founding members of the voltage-gated ion channel superfamily. The principal focus of this work is the investigation of the complement of potassium ion channels in our genome and its generalizabilty. An array of issues relevant to excitability is addressed, and a range of engagement in questions regarding the unity of life is proffered.

## INTRODUCTION

Potassium channels are an ancient protein family whose activities underlie life-sustaining processes, and whose polymorphisms and dysregulation are strongly linked with channelopathies. The fundamental pathways involving potassium channels are neuronal signaling, electrolyte balance, volume regulation, sensory signal transduction and immune response regulation. Their functional diversity is contributed in large part by various regulatory domains that modulate gating and conductance, and in eukaryotes, by their association with auxiliary components. Regulatory factors for potassium channels include membrane voltage, pH, cyclic nucleotides, intracellular calcium, ATP, polyamines, G-proteins, and temperature. The function of a potassium channel is determined by biochemical and electrophysiological characterization. The structure of the KcsA potassium channel realized historical notions of channel structure as a pore-forming unit in the plasma membrane that selectively passes ions in a high-throughput style (Doyle and others, 1998). Potassium channels possess a signature selectivity filter (almost always G[YF]G) (Heginbotham and others, 1994) in a pore region formed by flanking transmembrane helices. This structural motif constitutes the permeation pathway of the channel, and subfamily designations commonly indicate the attachment to this motif. The pore-forming monomers are referred to as α-subunits and usually assemble in the membrane as homotetramers. Protein sequence analysis has the potential to shed light on the diversity of potassium channels (Anfinsen, 1973). Hydropathy plots could infer transmembrane (TM) topology, which might be used to classify potassium channels. Analysis of TM topology was used by Jan and Jan (1997) to classify potassium channels in the animal kingdom into voltage-gated channels, which possess a 6TM topology, and inward-rectifying channels, which have a 2TM topology. Coetzee et al. (1999) expanded the scheme and included the 4TM topology class of two-pore "leak" potassium channels, first described by Lesage et al. (1996). In yeast, potassium channels with a 8TM-2 pore topology have been discovered (Ketchum and others, 1995). We were interested in a unifying principle of classification. We examined the feasibility of basing classification on the permeation pathway of potassium channels. Some properties that are individual to a subfamily may be intrinsic to the permeation pathway (for e.g., see Chen et al. (2002); Lippiat et al. (2000); Rich et al. (2002); Vergani et al. (1998)). A phylogenetic analysis of permeation pathways was a critical step in this effort.

## MATERIALS AND METHODS

### HUMAN POTASSIUM CHANNEL SUBFAMILIES:

The human genome has the most complex electrical signaling system. We adhered to the naming conventions of the HUGO Gene Nomenclature Committee (HGNC) guidelines. The HGNC database was searched for potassium channel subfamilies using the query "symbol begins with KCN" (KCN being a designated prefix). 15 subfamilies were returned, of which KCNE and KCNIP subfamilies are not pore-forming and therefore not retained. We enhanced the search with the following:

1. We used a less-specific query: "Full name contains potassium". In addition to the subfamilies obtained above, we identified another class that would be of interest to our analysis: the hyperpolarization- activated cyclic nucleotide-gated channels (HCN). We added these pacemaker ion channels to our list, since they are $K^+$- selective, and possess a transmembrane topology identical to voltage-gated K channels (Kaupp and Seifert, 2001).
2. The cyclic nucleotide-gated channels (CNG channels) are almost certainly homologous to potassium channels, since their transmembrane topology is identical to that of the potassium channel and there is good experimental evidence of close functional connection (Heginbotham and others, 1992). CNG channels are cation-selective and they are important in sensory processes of visual reception, auditory transduction and olfaction. The CNG channel family includes alpha and beta subunits and since both subunits are necessary to form a pore (Zhong and others, 2002), each subunit was treated as a separate subfamily (CNGA and CNGB subfamilies respectively).

| № | Subfamily Designation | Subfamily description (Common alias) | Primary function | Gating factor |
|---|---|---|---|---|
| 1. | KCNA | voltage-gated channel subfamily A (Shaker-related subunits) | Neuronal excitability | Membrane voltage |
| 2. | KCNB | voltage-gated channel subfamily B (Shab-related subunits) | Neuronal excitability | Membrane voltage |
| 3. | KCNC | voltage-gated channel subfamily C (Shaw-related subunits) | Neuronal excitability | Membrane voltage |
| 4. | KCND | voltage-gated channel subfamily D (Shal-related subunits) | Neuronal excitability | Membrane voltage |
| 5 | KCNS | voltage-gated channel subfamily S (‘modifier’ subunits) | Neuronal excitability | Membrane voltage |
| 6. | KCNF | voltage-gated channel subfamily F (‘silent’ modifiers) | Neuronal excitability | Membrane voltage |
| 7. | KCNG | voltage-gated channel subfamily G (‘silent’ modifiers) | Neuronal excitability | Membrane voltage |
| 8. | KCNH | human ether-a-go-go related channel (Herg) | Cardiac muscle excitability | Membrane voltage; cyclic nucleotide |
| 9. | KCNQ | KQT-like voltage-gated channel (KvLQT) | Cardiac muscle excitability | Membrane voltage |
| 10. | KCNMA | Large conductance calcium-activated channel (BK) | Skeletal muscle/Neuronal excitability | $[Ca^{2+}]_i$; membrane voltage |
| 11. | KCNN | intermediate/small conductance calcium-activated channel (IK/SK) | Skeletal muscle/Neuronal excitability | $[Ca^{2+}]_i$; membrane voltage |
| 12. | KCNJ | Inward-rectifying channel (IRK) | Maintenance of membrane resting potential | pH, G-proteins; ATP; temperature; polyamines |
| 13. | 1/2-KCNK | First pore of two-pore channels | “Leak” potential | pH, G-proteins; ATP; temperature; polyamines |
| 14. | 2/2-KCNK | Second pore of two-pore channels | “Leak” potential | pH, G-proteins; ATP; temperature; polyamines |
| 15. | HCN | Hyperpolarization-activated cyclic nucleotide-gated channels ($I_h$ channels) | Pacemaker activity | Membrane voltage; cyclic nucleotide |
| 16. | CNGA | Cyclic nucleotide-activated channel α subunit (CNG α ) | Sensory role | Cyclic nucleotide; membrane voltage |
| 17. | CNGB | CNGA- regulatory cyclic nucleotide activated β subunit (CNG β) | Sensory role | Cyclic nucleotide; membrane voltage |
| 18. | KCNSlack | Sequence like a calcium-activated K channel | Modulation of neuronal $[Ca^{2+}]_i$-activated K- channels (see Joiner et al. (1998) | $[Ca^{2+}]_i$; membrane voltage |

Table 1. Summary of pore-forming K channel subfamilies in the human genome. The primary source of information was the HGNC database. Note the addition of HCN and CNG subfamilies, and the splitting of the two-pore subfamily. The KCNSlack subfamily included here is discussed in the text later.

3. Each of the two pores in two-pore channels can form functional channels on their own (Saldana and others, 2002). Sequence searches suggested that two-pore $K^+$-channels do not have a prokaryotic counterpart (unpublished data). In anticipation of investigating the evolutionary origin of the two-pore channels, we considered the pores of the two-pore potassium channel as distinct, i.e., we assigned them separate subfamily definitions, 1/2-KCNK denoting the N-terminal pore-forming region, and 2/2-KCNK, the other one.

Table 1 includes these subfamilies along with their functional descriptors.

**COMPLEMENT OF HUMAN POTASSIUM CHANNELS:**

An Entrez (http://www.ncbi.nlm.nih.gov/Entrez/) text query was carried out for each major subfamily as defined by the above process to retrieve one representative member from the Entrez Genbank database. The sequence of this member was then used to seed PSI-BLAST (Altschul and others, 1997) searches that were iterated to convergence with non-stringent E-values (E-value < 10; inclusion threshold <0.01) to ensure the sensitivity of the search. We also carried out an independent text search on Entrez Protein with the following ternary query: "Homo sapiens [Organism] AND ((KCN* OR Kcn* OR kcn*) OR (Potassium* OR potassium*))". KCN is also the K Channel designation adopted by GenBank for annotating its potassium channels. This mixture of text-based searching and homology searching guaranteed that the sequence space of K channel in the human genome was covered. The human sequences mined from the above searches were consolidated. Each sequence in the resulting dataset was examined for serial satisfaction of the following:

1. the canonical selectivity filter triplet G[YF]G. Exceptions to take into account are the EYG, GLG and GLE variants of filters found in two-pore channels (unpublished observations) as well as the low-selectivity filters of CNG channels.
2. TM helices on either side of the selectivity filter. These are identified using hydropathy plots generated by the TMAP program (Persson and Argos, 1997), followed by manual inspection.

This ensured that the procedure is specific as well. We removed essentially duplicated sequences from the dataset (i.e., those with >=99% pair-wise sequence identity as determined by Li et al.'s algorithm (2001)). This process augmented the signal-to-noise ratio of the dataset. We obtained a set of 123 non-redundant potassium channel proteins, of which 23 consisted of two pore signatures in tandem.

**RESULTS AND DISCUSSION**

**MULTIPLE ALIGNMENT OF POTASSIUM CHANNELS:**

Potassium channels of different subfamilies do not yield to multiple alignment by standard methods. We used advanced techniques implemented in Clustal (Jeanmougin and others, 1998) such as:

1. sequence weighting
2. secondary-structure guided alignment (using the KcsA structure)
3. delaying addition of divergent sequences to the alignment

These techniques were still not successful in producing a satisfactory alignment of $K^+$ channels. We also used algorithms such as DbClustal and T-Coffee, and specialized matrices, such as transmembrane-specific substitution matrices (for e.g., PHAT (Ng and others, 2000)), but neither helped.

**Diversity of Potassium Channels:**

The source of this difficulty could be found in the diversity of potassium channels, which is manifested in many factors such as:

1. Gating factor
2. Membrane topology
3. Domain architecture
4. Character of tetramerization (homo- or hetero-tetramerization)
5. Association with auxiliary subunits
6. Mode of inactivation, and mechanism of channel closing
7. Electrophysiological characteristics (ionic conductance current, gating current)
8. Pharmacological profile (binding and sensitivity to blockers, toxins, etc.)
9. Kinetic considerations (rate of conformational conversions)
10. $K^+$-selectivity

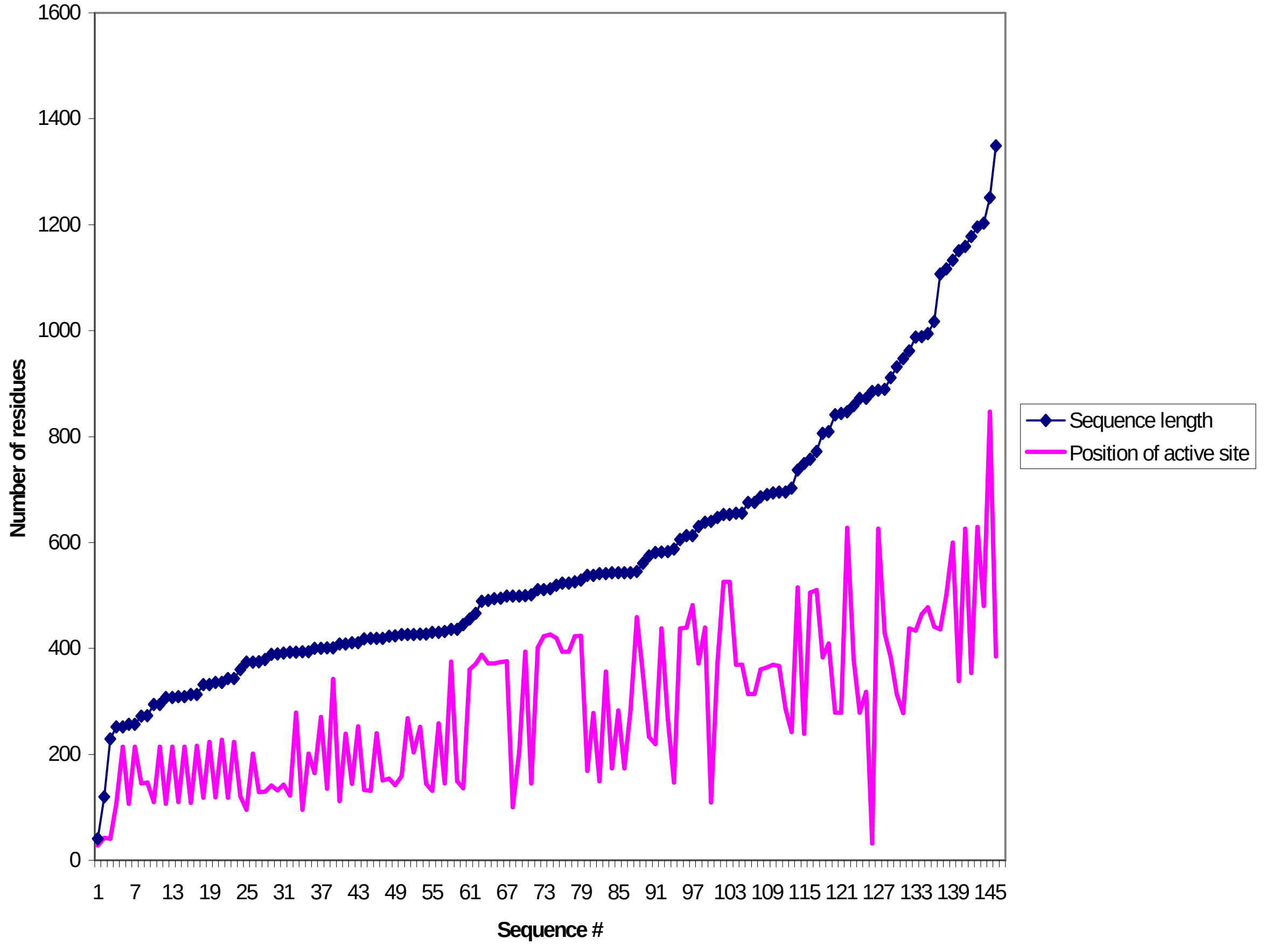

Figure 1A. A plot of the distributions of sequence lengths and active site positions, sorted by sequence length, for 146 $K^+$ channels. Sequence length is given by the upper tracing, and the location of the selectivity filter of the sequence is given by the lower tracing. The height of the lower line is the extent of the channel (in units of number of amino acid residues) from its N-terminus to the selectivity filter. The selectivity filter is defined as the position of the first glycine in the G[YF]G signature sequence. The figure illustrates the variation in the location of the channel selectivity filter.

For the discussion of all aspects of ion channels, consult Hille (2001). A plot of the location of the selectivity filter for each sequence relative to the sequence length shows large variations. (Fig. 1A). Fig. 1A revealed the following:

1. the range in the sequence lengths of potassium channels: from 120 residues to 1349 residues.
2. the uncertainty in the location of the selectivity filter relative to the rest of the sequence: it can be close to the N-terminus, close to the C-terminus, and almost any position in between.

The correlation of the location of the selectivity filter with the sequence length is very low, i.e. they are almost uncorrelated.

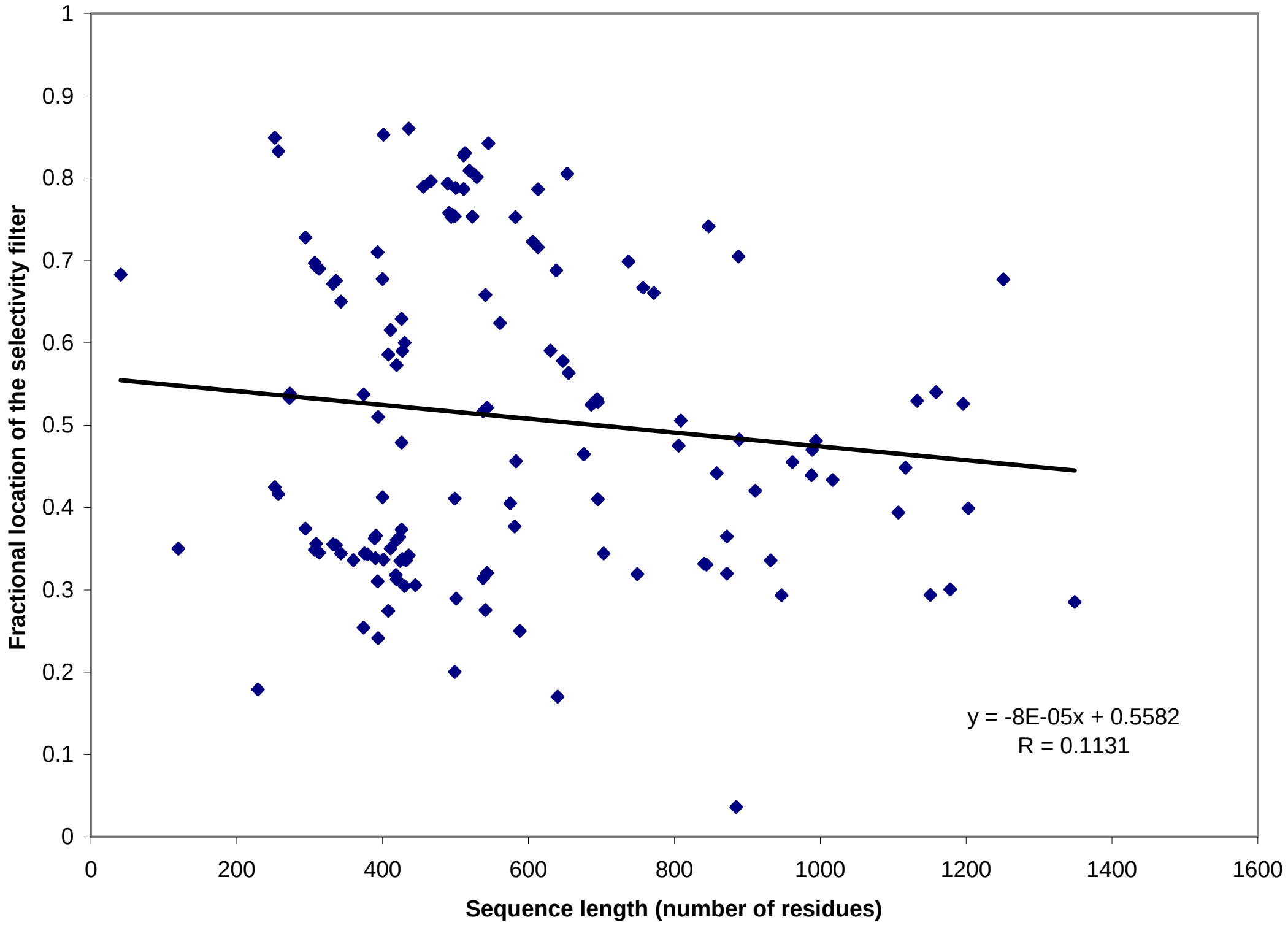

Figure 1B. Scatter plot of the fractional position (relative to sequence length) of the selectivity filter vs. sequence length. This plot implied that the location of the active site was not dependent on sequence length.

**Evolutionarily Conserved Features of Potassium Channels:**
To achieve the alignment of potassium channels, I started with the observation that the diversity of the $K^+$ channel family existed in a framework of evolutionary conservation. The selection for the selectivity filter might be exemplified by measuring the information content at a residue position in the alignment of many potassium channels. Since the residues of the selectivity filter are functionally important, change to the identity of these residues is disallowed. Some positions conserve more general physico-chemical characteristics. A measure of conservation, and the selection at a position, could be described by the Shannon entropy metric (Shannon, 1948), which is:

$$\text{Shannon's entropy } H = -\sum p_i \ln p_i \quad (1)$$

where the $p_i$'s are the frequencies of a residue type i (for e.g., i = Ala) relative to the total frequency at a given position, and the summation is over all residue types i at the given position.

Since information content at a position is defined as the decrease in uncertainty at that position, evolutionarily conserved regions would have lower entropy and variable regions would show maximum uncertainty. A plot of Shannon's entropy at each position along a sequence is termed an entropy plot. If the axis of the position is given in terms of the consensus residue of each position, the plot would reflect the strength of conservation at a position, along with the identity of consensus residue at that position. By juxtaposing indices of two types of information, namely conservation and composition, the figure would facilitate and simplify the assessment of functional contribution of notable positions and structures. This invention is called a "consensus-sequence entropy plot". Two consensus sequence entropy plots are shown below, one for the pore region (including the pore helix and the selectivity filter), and another for the S6/M2 inner helix of the permeation pathway.

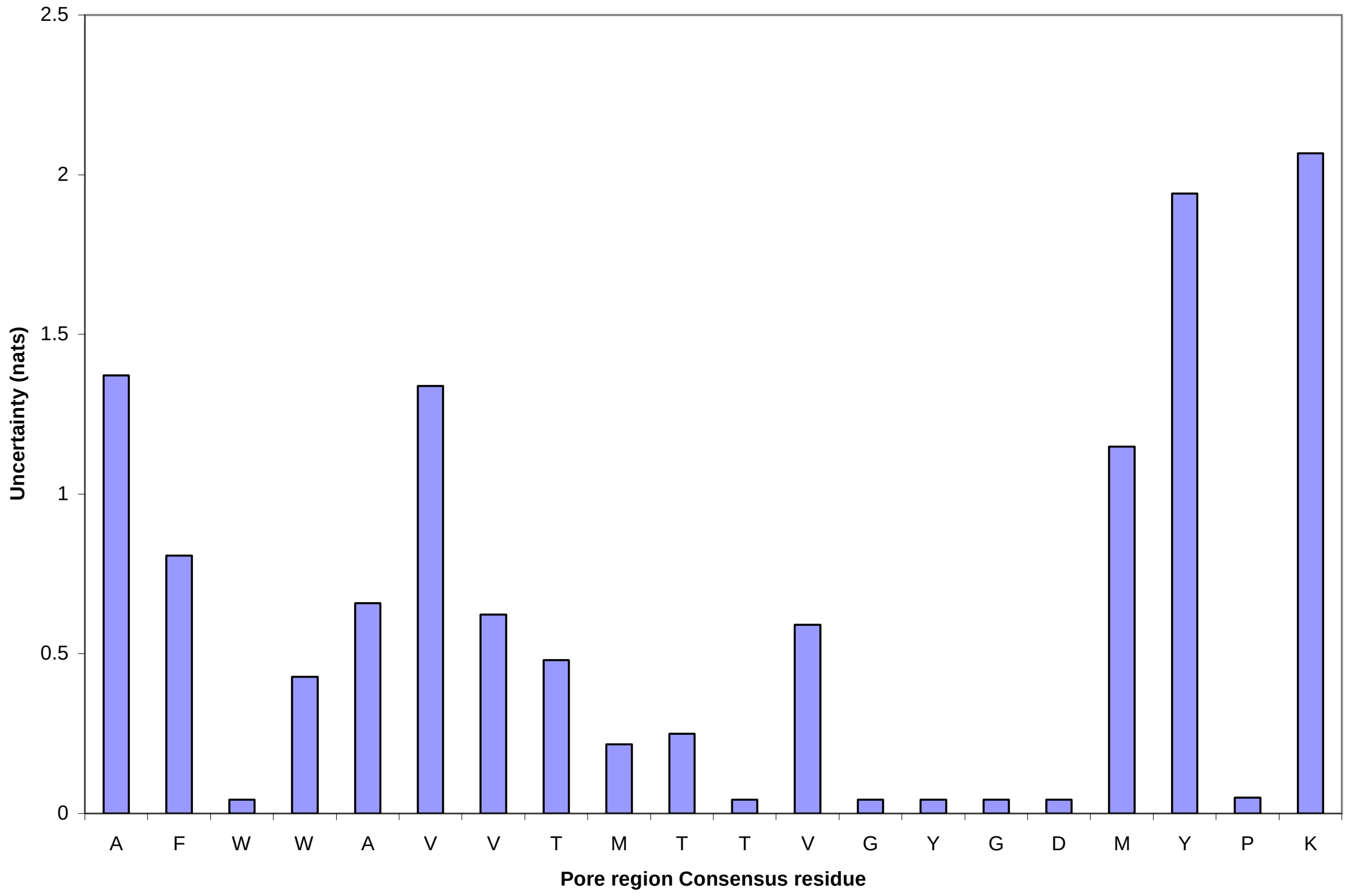

Figure 2. A consensus sequence entropy plot of the 20-residue pore region. The plot is constructed from the section of the pore region of a multisequence alignment of 90% non-redundant voltage-gated $K^+$ channels in all the major kingdoms of life, including eubacteria, archaea, protista, plants and animals. 147 sequences composed the alignment. The x-axis represents the consensus residue along the pore region. The height of the bars indicates the uncertainty in that position of the pore region. Note the minimal uncertainties in the positions of the active site GYG triplet, a Trp at the start of the pore helix, a Thr two residues before the active site, and a Pro following the active site.

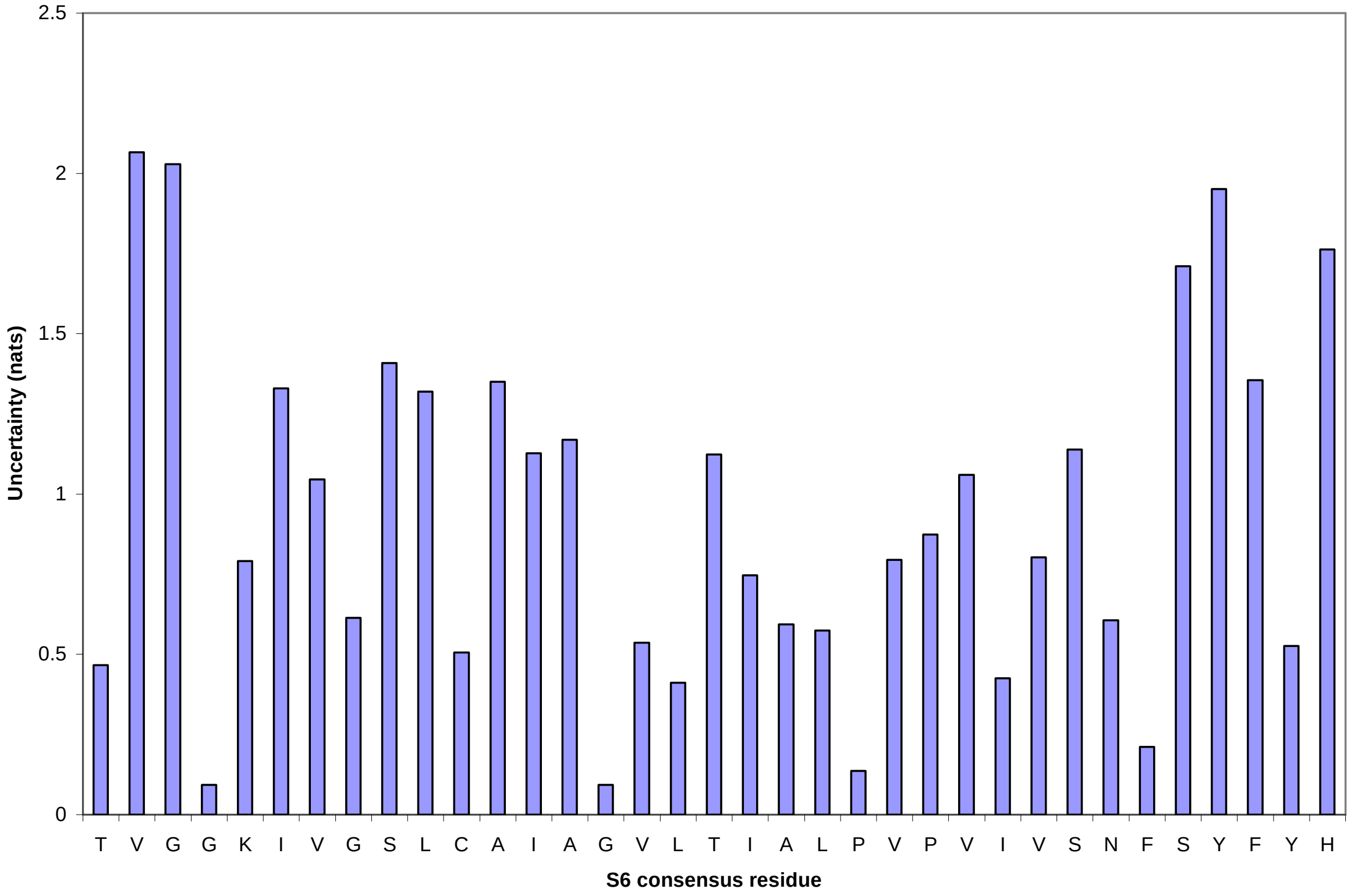

Figure 3. A consensus sequence entropy plot of a 34-residue region spanning the length of the S6 inner TM helix (extracellular to intracellular). The plot is constructed of the section of the S6 helix from a multisequence alignment of 90% non-redundant voltage-gated $K^+$ channels in all life, including eubacteria, archaea, protista, plants and animals. 147 sequences composed the alignment. The x-axis represents the consensus residue along the S6 helix. The height of the bars indicates the uncertainty in that position of the S6 helix. Notice that the glycine-gating hinge (located in the middle) has the least uncertainty among all the positions. The second, alternative gating hinge located at Ala five residues from the Gly-gate allows considerable uncertainty, perhaps implying that this position may not be very important in the gating of voltage-gated $K^+$ channels. The first Pro of the P-V-P motif that is a supposed motif for voltage-gating is nearly minimally uncertain, i.e. it is highly conserved in most of these channels (in all species). Also note the conservation of the Gly that is 11 residues above the gating hinge glycine.

It is clear that the alignment depended on the extraction of the permeation pathway of each pore in the dataset. The permeation pathway provided the following features to anchor an alignment:

1. the hydrophobic character of S5/M1
2. the aromatic cap of the pore helix
3. the constitution (including composition) of the pore helix
4. the signature selectivity filter
5. the hydrophobic character of S6/M2.

| Position | Kv | IRK | HCN | CNG α | CNG β |
|---|---|---|---|---|---|
| -10 | WYF | -[a] | FWY | FYW | FYW |
| -9 | - | WYF | - | - | - |
| -5 | TAISV | TAISV | H | TI | T |
| -2 | TSGQ | TSGQ | C | T | T |
| 1 | G | G | G | G | G |
| 2 | YFL | YFL | Y | ED | G |
| 3 | G | G | G | T [b] | L |

[a] '-' symbol denotes that the position is not significant for the respective category.
[b] valine (V) in the case of CNGα of C. elegans.

Table 2. We denote the absolutely conserved first glycine of the selectivity filter triplet as position 1. 'Kv' includes all voltage-sensitive subfamilies. The given patterns distinguished the categories.

The anchor points in the ion selection region for each subfamily are provided in Table 2.
To accommodate errors in the prediction of transmembrane helix ends, we extended sequences by four residues on either side of the permeation pathway, i.e, N-term to the outer helix, and C-term to the inner helix. Two permeation pathways were obtained for those channels with two selectivity filters. We generated a total of 146 (100 plus two for each of the 23 two-pore channels) permeation pathway sequences. The average length of the permeation pathways was 113 residues, which is about 20% the average length of a full channel. The 146 permeation pathways so obtained were aligned in the following three-tier process:
i. sequences of approximately the same lengths were aligned into a profile.
ii. the profiles were ordered in increasing length
iii. the profiles were sequentially joined through cycles of profile-profile alignment.

This heuristic was developed to minimize alignment errors in the turret region. Efforts in aligning $K^+$ channels gained the observation that the turret region was a source of the most errors in alignment, due to its high variability. This variability was natural for the region, since its function is to protect the channel from variant peptide toxins (Olivera, 1997). The overall alignment was then compared with other available alignments. For e.g., the alignment of cyclic nucleotide-gated channels (which have a deletion mutation in the selectivity filter region) was verified with the results of Leng et al. (2002), and Warmke and Ganetzky (1994). In some cases, structure-based manual alignment was done. The inner helix alignment was inspected for the alignment of the glycine hinge positions (Jiang and others, 2002), and the further intracellular bundle crossing. The patterns of outer helix, pore helix, selectivity filter, and inner helix in the comprehensive alignment are conserved hallmarks of potassium channel alignments, notably the previous alignment from our co- workers (Shealy and others, 2003).

The alignment is given in the Supplementary Information as Figure S1, and was used to generate the dendrogram shown in Figure 4. The dendrogram would function as a classifier of potassium channels if there exist nodes on the dendrogram that radiate each subfamily. Ideally, there must be distinct clusters for each subfamily. The clusters formed by 1/2-KCNK and 2/2-KCNK subfamilies are readily noted. A high degree of divergence is observed within the 2/2-KCNK subfamily, cerating two clusters. The one-to-one correspondence between the clusters of the dendrogram and the potassium channel subfamilies defined in Table 1 are noted on the right hand side of Figure 4. Many sequences in our dataset were missing annotation. This situation was rectified by performing full-sequence searches to validate the subfamily of these sequences as determined by the dendrogram. For example, if the permeation pathway of a given sequence was placed in a cluster of known calcium-activated channels, we verified that the full sequence of the unannotated sequence had calcium-binding domains. In the course of this investigation, we noted a subfamily of Slack (Sequence like a calcium- activated K channel) channels in the human genome. Cloned rat Slack channels were intermediate-conductance $K^+$ channels that obligatorily heterotetramerized with the α-subunits of other $[Ca^{2+}]$i-activated $K^+$ channel subunits (Joiner and others, 1998). Existence of two Slack channels in the human genome was previously reported (Yuan and others, 2003). In our computational study here, we obtained three putative Slack channels. The Genbank records for each of these channels however lacked annotation. Each of these sequences was most closely related to the rat Slack $K^+$ channel subfamily.

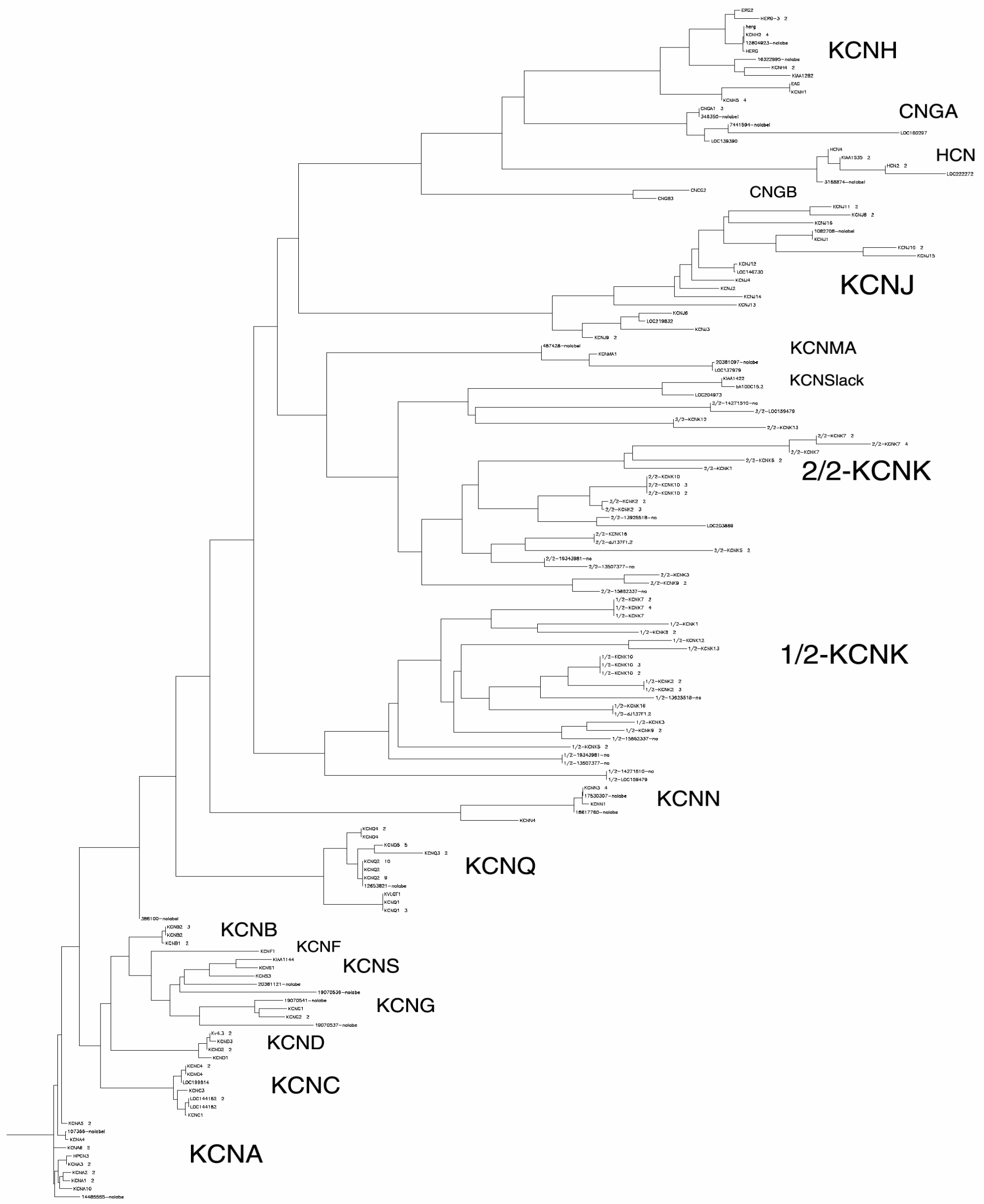

Figure 4. Dendrogram representing the pattern of clustering of permeation pathways of human K+ channels. Subfamily designations from Table 1 are used to label the clusters, the size of the label being in approximate proportion with the size of the cluster. The dendrogram was constructed by applying the neighbor-joining algorithm on the computed distance matrix of the alignment. The phylogenetic utilities in ClustalW were used for computing the matrix and the dendrogram. NJPlot (Perriere and Gouy, 1996) was used for visualizing the dendrogram.

The functional annotation and classification of the potassium channels in the human genome are given in (Palaniappan, 2005). To push for a consensus in the ion channel community about the number of $K^+$ channels in the human genome, we present the following comparisons. There are 7 CNG channels in our set; only 6 were noted in (Kaupp and Seifert, 2002). Of Kv channels, 35 are identified in this work, comparable numbers being 22 given in (Yellen, 2002) and 26 in (Ottschytsch and others, 2002). The total number of $K^+$ channels identified in our analysis is 123. A study by Yu and Catterall (2004) identified an 88-member human genome-complement of $K^+$ channels. Gulbis and Doyle (2004) placed the number of $K^+$ channel genes in the hundreds. Alternatively spliced gene products are likely to increase the number of $K^+$ channels expressed by the human genome.

The phenomenon of co-evolution is responsible for the ability of the permeation pathway to separate potassium channels into their respective subfamilies. It is studied in (Palaniappan, 2007). Here, we present an alternative method for asserting co-evolution between the catalytic and regulatory domains of potassium channels. We derived the consensus sequences of the permeation pathway and the voltage sensor, and compared the correlation of divergences of an extant permeation pathway from its consensus sequence and its corresponding voltage sensor from *its* consensus sequence. In essence, we were modeling the evolutionary process as a random walk of the individual sequences from a consensus sequence located at the origin. We obtained a correlation coefficient r = 0.4631 for 137 pairs of permeation pathway distances and voltage sensor distances. The significance of the correlation was estimated for 1000 bootstrap runs. The result could be represented in terms of a z-score, which was $1.030*10^5$ for this calculation; the corresponding p-value is 0.0000. Therefore, classification based on the evolutionary similarity in the structural scaffold of the active site is based on a biological principle. Our results discredited a notion of functional modularity of domains.

**LINEAGE-SPECIFIC EXPANSIONS:**

All voltage-gated subfamilies involved in electrical excitability (namely the KCN A, B, C, D, F, G and S subfamilies) formed neighboring clusters in the dendrogram. A similar co-clustering is observed for subfamilies that possess the cyclic nucleotide-binding domain. To obtain clarity in these studies, we reconstructed the phylogeny of the human $K^+$ channel subfamilies using the consensus sequences of each subfamily. The construction of the multiple alignment of the consensus sequences of the various subfamilies is described in (Palaniappan, 2006). Based on this alignment, we constructed a neighbor-joining tree and bootstrapped the resultant tree. This tree is shown in fig. 5. The root is assumed to be equidistant from all the tips. The following observations were made from the tree:

1. The voltage-gated $K^+$ channel subfamilies, viz. KCNA, KCNB, KCNC, KCND, KCNF, KCNG and KCNS, all clustered together into one big group. The bootstrap support for this group is nearly 100%. This overwhelming evidence for the monophyly of the voltage-gated $K^+$-channel family implied that the acquisition of the voltage sensor happened only once, after which the voltage-gated family has been diverging from the other ion channel families.
2. The class of $K^+$ channels that possesses a cyclic-nucleotide binding domain (CNBD) comprise four subfamilies: the hERG family, the cyclic nucleotide-gated a and ß subfamilies, and the hyperpolarization-activated $K^+$ channel. The hERG channels and HCN channels are sensitive to both membrane voltage and cyclic-nucleotides. The dendrogram provided good support (bootstrapped confidence >68.8%) for the monophyletic grouping of the CNBD-containing channels. The acquisition of the cyclic-nucleotide binding domain provided the stimulus for the functional divergence of this group of channels from a voltage-gated $K^+$ channel ancestor.

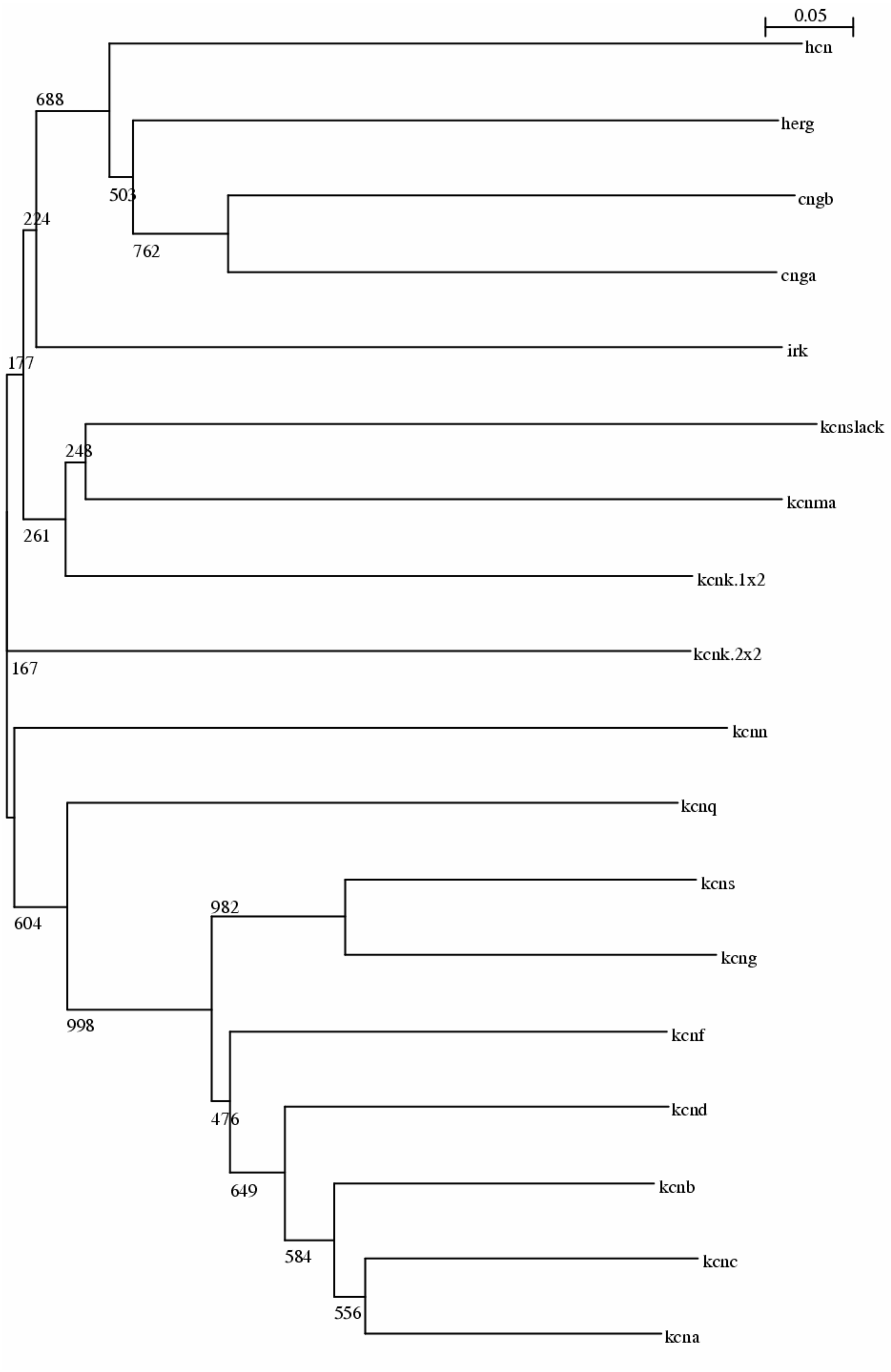

Figure 5. A neighbor-joining tree from the consensus sequence multiple alignment. 1000 bootstrap runs were made and the bootstrapped values of each node are given adjacent to it. Note the substantial statistical support for the two major lineages in the tree: one comprising the large voltage-gated class of $K^+$ channels (bootstrap support > 99%); the other, that possess a cyclic-nucleotide binding domain (bootstrap support > 68%). For the subfamily shorthand used here, refer to Table 1. The tree was drawn using NEIGHBOR of PHYLIP and bootstrapped using ClustalW.

3. A common ancestor for KCNQ and voltage-gated $K^+$ channel family is well-supported. No clear conclusions could be drawn of the origin of the calcium-activated $K^+$ channels represented by KCNMA, KCNN and KCNSlack. The small-conductance calcium- activated channel appeared to be most closely related to the voltage-gated class, however the statistical support is not considerable. The channels unresponsive to changes in membrane voltage cluster independently of the voltage-gated $K^+$ channel family on the dendrogram. They include the large family of KCNJ channels, which are gated by any of multiple factors excluding voltage, and the two-pore KCNK channels.

**Origin of the Two-Pore Channel Subfamily:**

We investigated the weight of evidence on the origin of the two-pore channels. The multiple alignment of the human two-pores was extracted as a subset of the genomic alignment of $K^+$-channels (see pp. 65-67 in (Palaniappan, 2005)). This alignment provided the basis for the discussions below.

1. Selectivity filter: the amino acid identity of the middle residue of the selectivity filter triplet in the first and the second pores are consistently different. In human, the Gly's of the selectivity filter are absolutely conserved; the middle position tolerates three different amino acid residues, namely Tyr, Phe, and Leu. The identity of the middle position is physiologically significant. We find that the middle residue of first pores is predominantly Tyr, whereas that of second pores is predominantly Leu. The counts are given in Table 3.

| Subfamily: | Tyr | Phe | Leu |
|---|---|---|---|
| 1/2-KCNK | 19 | 4 | 0 |
| 2/2-KCNK | 0 | 18 | 5 |

Table 3. Count of the identity of the middle residue of the selectivity filter in the first pore and second pore of 23 two-pore channels of human.

2. Turret region: The turret regions of the 1/2-KCNK subfamily are consistently much longer than the turret regions of the 2/2-KCNK subfamily. The shortest turret in the first pore is 49 residues, and the longest turret of the second pore is only 6 residues.

3. Distance matrix analysis: We analyzed the distance matrix of all the first pore and the second pore sequences together. The closest second pore relative of a first pore sequence is still more distant than all the other first pores, and vice versa.

4. Phylogeny and Bootstrap analysis: Based on the distance matrix obtained, we constructed a neighbor-joining tree of the first and second pores of two-pore channels. This tree is shown in Figure 6. The 1/2-KCNK and 2/2-KCNK subfamilies were represented in the tree by exclusively monophyletic clusters. We found that there was greater than 50% bootstrap significance for the node that represented the branching point of the first pore and the second pore clusters.

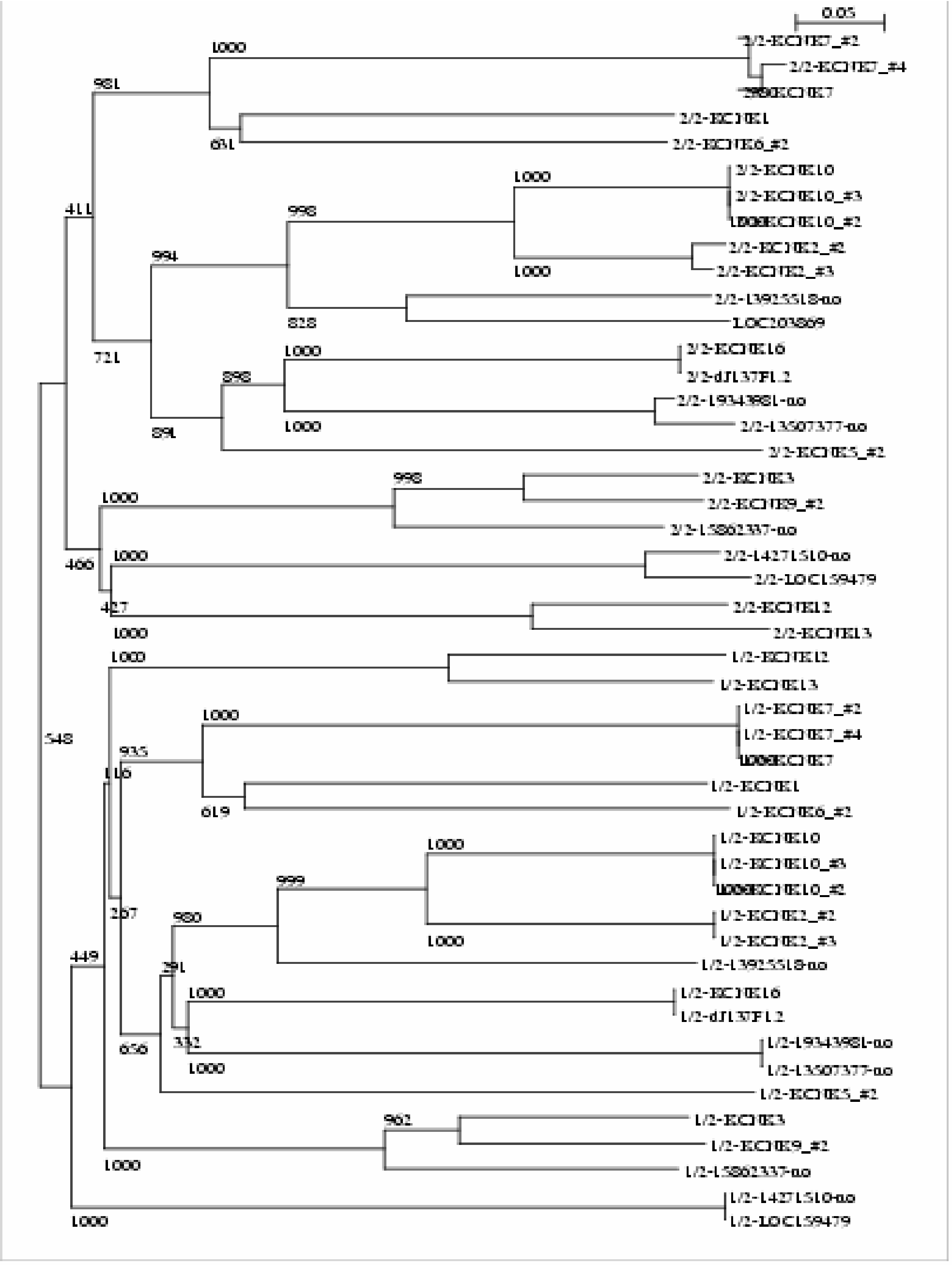

Figure 6. An NJ tree of the individual pores of the human two-pore complement. The first pores are prefixed with 1/2, and the second pores are prefixed with 2/2. It is seen that the first pores, and the second pores form exclusively monophyletic clusters. The replication of each node in 1000 bootstrapped runs is given. The replication of the connection between the two clusters is 54.8%.

5. Evidence for concerted evolution: The methodology for co-evolutionary analysis described in Chap.3 is applied to the analysis of co-evolution between the first pore and the second pore. We found r = 0.86 for the co-evolution between the two pores of the same channel. The significance of the coefficient of correlation was estimated similarly; r was highly significant (z-score = 1.6x105; p-value = 0.000000). Thus the two individual pores have been co-evolving for a long time, and this co-evolution must be accounted for by a hypothesis of the origin of the two-pore channel.

Two hypotheses are consistent with the sum of the above observations on the issue of the origin of the two-pore channels. The first one is: there was tandem duplication of a single-pore $K^+$ channel gene, followed by a single base mutation in the selectivity filter code of one of the repeats, along with the substitution of a long turret in the other repeat, followed by concerted evolution. This hypothesis demanded a degree of evolutionary stability of the intermediate stages, and since not one of these intermediate stages is extant, the hypothesis was wanting.

The second hypothesis is that there was a fusion of two distinct $K^+$-channel ancestors, one of which became the first pore, and the other the second pore. The first ancestor had a selectivity filter with Tyr, and a short turret region; the second ancestor had a selectivity filter with Phe, and a much longer turret region. The plausibility of both ancestors exists. The fusion event was followed by concerted evolution. This hypothesis explained all the data, and by virtue of its parsimony, we concluded that the two-pore channels descended from a fusion event of two distinctive $K^+$ channel ancestors.

Our investigations into the origin of the two-pore channels led to a counter-intuitive conclusion and underscored the domain of applicability of the empirical method. The hypothesis advanced to account for the patterns of evolution of sodium and calcium channels, namely twice over internal duplication, automatically suggested that the two-pore channels similarly evolved. Our investigation established that the two-pore channels emerged from an independent event (of gene fusion) in the course of eukaryotic evolution.

**SUBFAMILY-WISE DISTRIBUTION OF IONIZABLE RESIDUES IN PORE:**
In this section, we explore a hypothesis that ion channel pores tend to be lined by 'rings' of ionizable residues (especially acidic residues), which enhanced the permeation process. The count of ionizable residues from the selectivity filter to the first half of the inner helix for each subfamily was obtained from the alignment of consensus sequences in (Palaniappan, 2006).

| № | # ionisable residues (acidic +basic) per subunit | # acidic residues per subunit | subfamily distribution | # subfamilies |
|---|---|---|---|---|
| 1. | 1 | 1 | KCNC | 1 |
| 2. | 2 | 1 | KCNA, KCNG, KCNK.1, HCN, KCNN, KCNSlack | 6 |
| 3. | 2 | 2 | CNGB | 1 |
| 4. | 3 | 1 | KCNB, KCND, KCNF, KCNQ, KCNK.2, KCNMA | 6 |
| 5. | 3 | 2 | KCNS, KCNH | 2 |
| 6. | 4 | 3 | KCNJ | 1 |
| 7. | 5 | 4 | CNGA | 1 |

Table 4. Count of ionizable residues in a single subunit of various $K^+$ channel subfamilies.

From the data in Table 4, we see that every subfamily has at least one ionizable residue per channel monomer. It is recalled that CNG-α subunits obligatorily heterotetramerize with CNG-ß subunits in a 3:1 stoichiometry, as well as the fact that KCNK.1 and KCNK.2 are combined in polypeptide sequence into a single dimer. Weighting the number of ionizable residues by the stoichiometry of the fully assembled channel, the total number of ionizable residues per channel tetramer is calculated using columns 2 and 5 of Table 4. [(1*1 + 2*5.5 + 2*0.5 + 3*5.5 + 3*2 + 4 + 5*1.5)*4/18 = 10.4.] The ~10 ionizable residues in the permeation pathway of the average $K^+$ channel tetramer could participate in creating rings of ionizable residues that promote the conduction of $K^+$ ions.

**STRUCTURAL CONSERVATION OF THE PERMEATION PATHWAY:**

There was evidence that the structure of the permeation pathway of the prokaryotic potassium channel, KcsA, is conserved in eukaryotes (LeMasurier et al., 2001; Lu et al., 2001). A toxin-binding study of the turret outer vestibule of a prokaryotic channel and a eukaryotic channel provided evidence for the view that there is significant structural conservation of channel structure across life (MacKinnon and others, 1998). Here, we explore the conservation of all regions of the permeation pathway using computational tools. We aligned the KcsA potassium channel onto the alignment constructed of the human channels, and noted the conservation of amino acid character for each residue position when it is mapped onto the KcsA structure. For assessing the conservation in prokaryotes, a similar alignment and calculation of conservation of the permeation pathway of prokaryotic potassium channels was done. (Sequences were retrieved using a Psi-Blast analysis of the KcsA prokaryotic channel sequence.) A numerical score of conservation was computed using ConSurf (Armon and others, 2001), an algorithm based on phylogeny information. The two sets of conservation scores, one for human and the other for prokaryotic potassium permeation pathways, were projected on the KcsA structure, with the gradation in colour reflecting the gradation in conservation (Figs. 7 and 8).

Remarkably, the conservation patterns in the prokaryotic and human channel permeation pathways are highly similar. The conservation in the human sequences is certainly stronger, indicating that diversification into subfamilies has tended to conserve more strongly than the independent divergence of channels in prokaryotes.

In both human and prokaryotic sequences, there is strong conservation in the pore helix and selectivity filter and in those regions of the TM helices involved in intrasubunit packing; i.e., packing between the inner helix and the pore helix and packing between the inner and outer helix. These conservation patterns suggest that essentially the same packing pattern has been preserved for the majority of potassium channels throughout all of evolutionary history since the first potassium channels in prokaryotes. There is also strong conservation along the lumenal face of the inner helix, suggesting that the detailed structure of the intracellular end of the permeation pathway is conserved to perhaps nearly the same degree as the pore helix and selectivity filter that define the extracellular end of the pathway. The conserved glycine in the middle of the inner helix postulated as the gating-hinge is very highly conserved in the figure. The moderate conservation of the interface between the inner and outer helices suggested the partial conservation of the mechanical coupling between these helices in a variety of $K^+$ channel subfamilies.

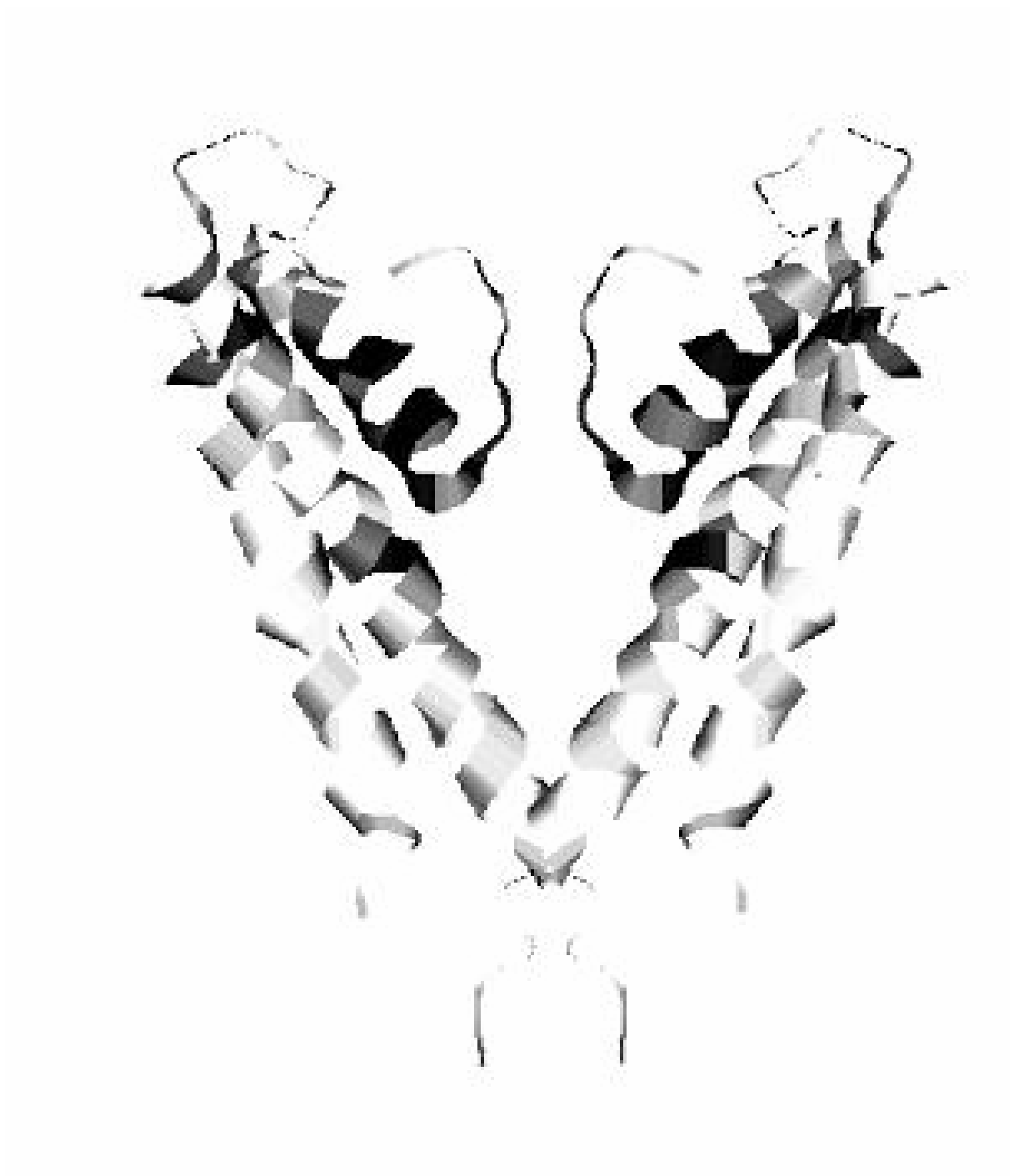

Figure 7. Structural conservation map of human potassium channels. Conservation grades were computed using the Consurf algorithm and the picture was visualized using Rasmol (Sayle and Milner-White, 1995) by replacing the B-factor in the PDB structure file with the conservation grades. More pronounced regions are more conserved. Only two diametric monomers are shown for the sake of clarity.

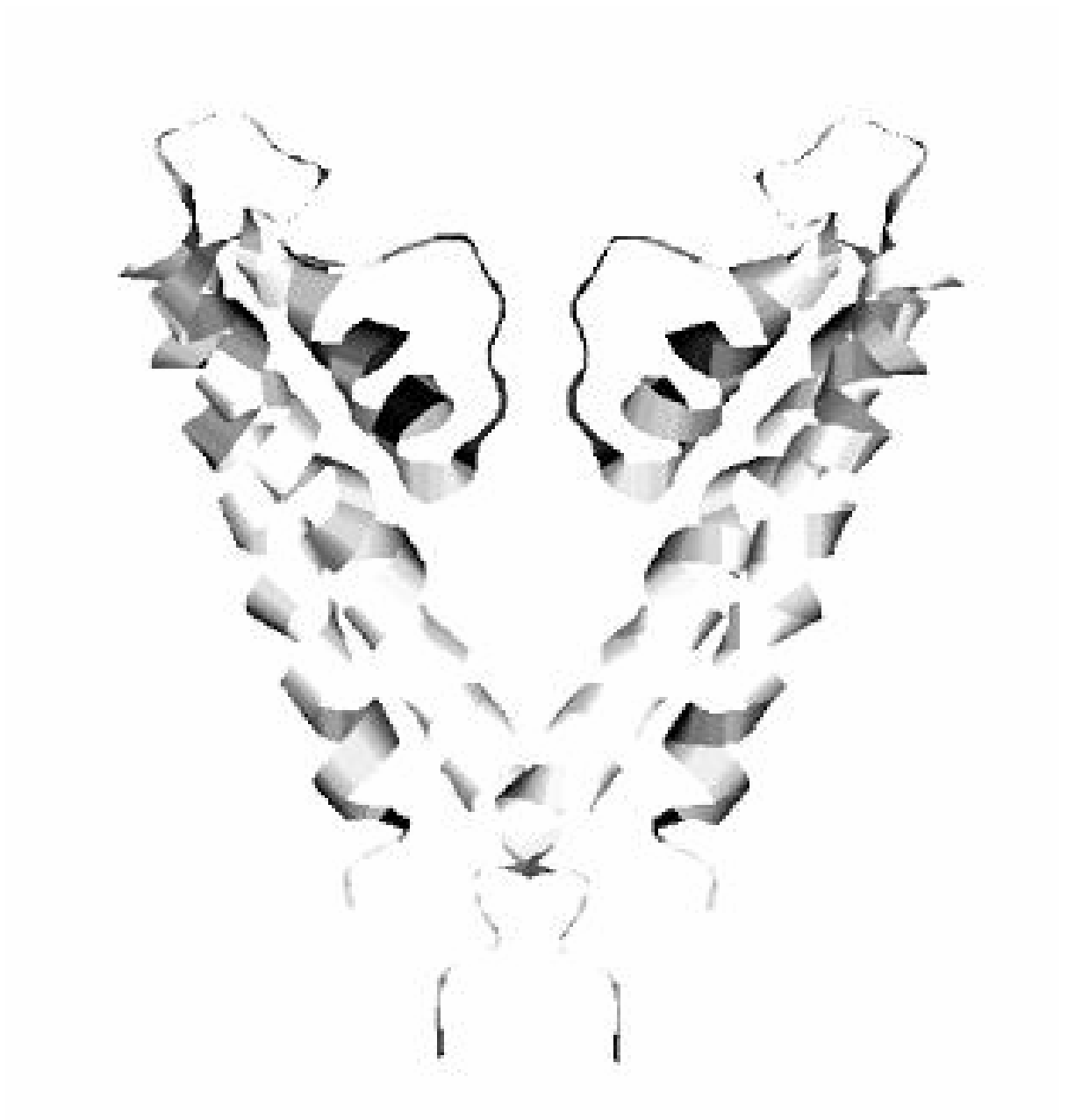

Figure 8. Structural Conservation map of prokaryotic potassium channels. Conservation grades were computed in a manner similar to Fig. 7. Note the pattern of similarity between the maps.

The regions not strongly conserved are regions of intersubunit packing, where the different monomers that make up the tetrameric potassium channel pack up with each other. We note that the vast majority of potassium channels are homotetramers. The variable surfaces at the point of intersubunit packing would mitigate against the assembly of heterotetramers. Further, the variability in these regions is consistent with the specificity necessary for channel regulation.

**VISUALIZING LARGE PHYLOGENIES:**
We were interested in using visualization technology to aid the analysis of mammoth, humongous phylogenies. Similarity relationships among biological entities are often conceptualised and visualized as reticulate, branching trees. Branch points reflect evolutionary events of divergence whereby an ancestor diverged into two classes of progeny. The visualization of trees has been constrained to small trees. This offered lower confidence to the inferences from the tree, as well as constrained the range of data that could be employed. However, with the exponential growth in the availability of protein sequences, it is theoretically possible to construct phylogenies that represent the detailed evolution of a protein family.

There are two components to understanding the evolution of a protein family. One is to understand the time of divergence and split of the protein family in various species. The second is to understand the duplication and diversification of the protein family into specialized subfamilies within each species. The first involves analysis of broad patterns, and the second involves analysis of the fine details of the phylogeny. It is advantageous to be able to do the two evolutionary studies in a concerted manner. I explored the total picture of a protein family, i.e., its origins and its lineage-specific expansions, in the context of $K^+$ channel proteins.

In partnership with the NCSA (National Center for Supercomputing Applications, University of Illinois) Scientific Visualization Group, we used the NCSA Wall Tile Display for large-scale phylogeny visualization. For a complex tree with many entities, regular computer or paper displays are not capable of simultaneously showing the fine detail and the overall pattern of relationships. Potassium channels, being both ubiquitous and very diverse, would be ideally suited to examining the efficacy of the wall tile display in exploring phylogenetic issues. A highly detailed $K^+$ channel dendrogram of many taxa was constructed. This tree, projected on an area and resolution twenty times a 17" desktop monitor, displayed all known prokaryotic (bacterial and archaeal) potassium channels and all the potassium channels from two animals, human and the nematode C. elegans. The animal channels were seen to separate into three classes depending on the type of gating involved: voltage-gated, ligand-gated, and "other" (for example, pH- and temperature-sensitive). For the ligand-gated and voltage-gated classes, prokaryotic channels were mingled with the animal channels; this indicated that these classes originated with prokarya. For the third class of channels, no prokaryotes were seen to cluster with the eukaryotic representatives, thus indicating a eukaryotic and more recent origin for this class compared with the other two classes. Another interesting aspect of ion channel evolution uncovered by the tree pertained to the origin of other ion channel families, the sodium ion-channel and the calcium ion-channel families in particular, and these were seen as arising from a particular region of the ligand-gated class of potassium-ion channels.

**CO-OCCURRENCE ANALYSIS OF ß-SUBUNITS OF $K^+$ CHANNELS:**
The integral membrane (α-) subunits of many voltage-gated ion channels are associated with an additional protein called the ß-subunit; in the case of voltage-gated $K^+$ channels, these are denoted as Kv ß. The fourfold rotational symmetry of the α- subunit continues in this subunit too and the Kv ß is a tetramer of oxidoreductase proteins. The structure of the cytoplasmic assembly of α-ß subunits reveals the ß subunit interacting non-covalently with the T1 domain of the α-subunit and forming a fourfold symmetric T1(4)ß(4) complex. The T1 domain is the tetramerization domain of voltage-gated $K^+$ channels and is entirely cytoplasmic. The presence of the cofactor NADPH in the active site of Kv ß proteins bears implications for the regulation of their activity. However, the precise biological function of these auxiliary subunits is not known yet. Their shared sequence homology with aldo-reductases and the presence of the nicotinamide cofactor strongly suggests a central energetic coupling role of the T1(4)ß(4) complex in the cellular redox regulation of the channel. In fact, the proximity of the NADPH-binding site with a Kv a subunit lends shape to a hypothesis about how the S4 helix, primarily governing cell excitability, can be directly coupled through structural changes in T1 to the enzymatic activities of the Kv ß protein. The genus distribution of the ß-subunits also is not clear with relation to the distribution of the α-subunit. For e.g., is co- occurrence of the ß subunit a must within the voltage-gated class of $K^+$ channels? Specifically, does the ß subunit co-occur with T1 domains? The answer to this question is not clear. In fact, there are bacterial genomes with sequences homologous to the ß subunit, yet no sequence codes for a T1 domain in the same genome. Some Kv ß proteins have an N-terminal sequence, with homology to the α-subunit N-terminal inactivation ball that can induce rapid channel inactivation. ß subunits, in addition, increase the efficiency of cell-surface expression of Kv α-subunits.

A very interesting question arises from the consideration of the cycle of regulation. Could it be possible that, in addition to the ß subunit being responsible for rapid stimulus-dependent channel inactivation, the channel itself regulates the activity of this oxidoreductase enzyme in response to cellular redox potential demand? It was possible to determine the ß subunit residues that interacted with the residues of the T1 domain from the crystal structure of the ß subunit-T1 assembly complex (Gulbis and others, 2000). Interacting residues are within non-covalent bonding distance. The databases of protein families/protein motifs (Pfam & PROSITE) indicate that ß subunits belong to the larger class of NAD-binding oxidoreductases. The motifs derived for the larger family did not include of the regions of interaction of the ß subunit with the T1 domain. The T1-interaction surface of the ß subunit defines its role in the context of channel function, and thus we concluded that the region of interaction would distinguish the ß subunits from other NAD-binding oxidoreductases. We extended this analysis of the signature of the ß subunit with Blast searches. We used as query sequence the ß subunit whose structure is known (PDB ID: 1EXB). We analyzed the patterns of the E-values of the hits in relation to the degree of conservation of the residues important in functional interaction with the T1 domain. There are three ranges of E-values. The first range comprises the true positives, i.e., homologous ß subunits, and these reflect near-total conservation of the functional interface of interest. There is huge gap in sequence space before the next hit is obtained. There is only 20% or lower conservation of the residues of the functional interface in the second range. These sequences must be belonging to the same larger family; they are perhaps other classes of NAD-binding oxidoreductases. They may not be effective ß subunits due to the absence of selection pressure on the interaction surface. As we moved to the third range of E-values, the hits are nearly all false positives, and there is little conservation in the functional regions. We inferred that the border between the first range and the second range of hits defined the boundary of the ß subunit subclass of NAD-binding oxidoreductases. We also observed that plant ß subunits tended to deviate from the pattern of conservation seen in the ß subunits of the animal kingdom, perhaps implying that plant ß subunits used an alternative surface of interaction.

Our earlier questions gain importance in the context of the prokaryota. Of the prokaryotic hits to the ß subunit query sequence, none showed any appreciable conservation in the region of the interaction surface. We also investigated the occurrence of T1 domains in prokaryotic voltage-gated $K^+$ channels, and did not find evidence for the existence of a T1 domain in any prokaryotic channel. The implication of these complementary analyses is significant. It bears relevance to the nature of functional association between channel α- and ß-subunits in prokaryotes. It also raises fundamental questions about the role of ß subunits in prokaryotes. A corresponding analysis of T1 domains in α-subunits turned up a smooth progression of E-values, and thus was not valuable in inference studies.

**COMMON ANCESTOR OF THE VOLTAGE-GATED ION CHANNEL SUPERFAMILY:**
The three primary classes of voltage-gated ion channels are $K^+$-, $Na^+$-, and $Ca^{2+}$-channels. $K^+$ channels were originally 2TM channels, which then evolved voltage-dependence through the acquisition of a 4TM voltage sensing module, thus giving rise to a 6TM Kv channel. An argument for the evolution of voltage-gated $K^+$ channels into sodium and calcium channels, which are invariably voltage-gated, is given below. It provides a prototypical example for an original technique for exploring evolutionary origins among distant families that are likely to have evolved in an unknown sequence. The technique is based on the construction of consensus sequences from knowledge-based aggregation of multiple data sets. We could consider that the group of voltage-gated ion channels forms a superfamily, defined by the characteristics of voltage-dependence, and functional similarity.

All prokaryotic organisms possess mechanisms for conduction of potassium ions. Some possess $K^+$ transporters; many prokaryotes possess a $K^+$ channel that is unmistakably homologous to eukaryotic $K^+$ channels. $K^+$ channels in eubacteria and archaea possess two transmembrane architectures, 2TM and 6TM, which correspond with the eukaryotic inward rectifying and voltage-gated $K^+$ channels, respectively. A 4-TM two-pore $K^+$ channel has not been observed in the prokaryotes. $K^+$ channels are coded by the genomes of most prokaryotes whose sequencing is complete, and their ancestral origins may lie in the earliest form of life.

Most prokaryotes lack channels that conduct sodium and calcium ions, and active mechanisms are necessary to mediate the transport of these ions. Notable exceptions exist. These few prokaryotic species that possess members of the sodium and calcium family are invaluable in phylogenetic studies. An ‘outlying’ gene in the primitive forms of life is often a determinant of the origins of a gene family. (Exceptions to this rule arise from the mechanism of lateral transfer. A gene could be integrated in a prokaryotic genome through a form of lateral transfer from a eukaryotic host, in which case the gene is not a candidate for a ancestral entity.) We are interested in the precursors to the sodium and calcium ion channel family. The diversification of precursors is achieved through evolution. In the study of the precursor to sodium- and calcium-channels, we must first test the existence of members of these families in prokaryota. If the existence test is positive, we must then examine if these putative precursors could have arisen through lateral transfer. If it could be established that these members are intrinsic prokaryotic genes, i.e. they were not laterally transferred, they could be used for phylogenetic studies.

Sequence searches were performed to uncover members of the sodium- and calcium-family in prokaryota. We found two sodium channels and one calcium channel in prokaryota. The sodium channels were from alkaliphilic bacteria, which live in an alkaline habitat. By virtue of their extreme lifestyle, lateral transfer can be ruled out as an origin of their genes. The calcium channel is from the Paracoccus sp., a division of proteobacteria. The single 6TM region of each of these sequences was used. The paracoccus channel bore homology to the second and third repeats of a calcium channel in C. elegans and limited homology to the first and fourth repeats of the same sequence, confirming its ion selection property. We found that the II and III repeats of calcium channels were closer than the I and IV repeats and vice versa. The same observation was true of sodium channels (both data not shown).

Let us consider the postulate that potassium channels are the prototypical members of the voltage-gated ion channel superfamily. The sequences of sodium and calcium channels are similar four-fold repeats of the sequence of a $K^+$ channel. A parsimonious hypothesis would suppose that sodium and calcium channels evolved over two rounds of duplication of a precursor $K^+$ channel. This would recognize $K^+$ channels as the founding members of the voltage-gated ion channel superfamily.

6TM $K^+$ channels were obtained from a proteobacterium, and an archaeon. These were from Bradyrhizobium japonicum and Methanosarcina, respectively. Consensus sequences represent a reconstruction of the probable ancestral sequence of a set of sequences, usually assuming that all the members were evolving at the same rate for the same period of time. We constructed the following three types of consensus sequences to study the evolutionary history of the voltage-gated ion channel superfamily:

1. The primeval $K^+$ channel represented by the consensus sequence of the two $K^+$ channels selected for the analysis.
2. The ancestor of calcium and/or sodium channel: this is theoretically responsible for the divergence between sodium and calcium channels. Note that the ancestor represents a descent from one of the channels which functionally diverged to create the other channel. It is represented by the consensus sequence of the three prokaryotic members of the sodium and calcium channel families identified above.

3. We were interested in the ancestor that split away from the $K^+$-channel family and gave birth to the calcium and sodium channels. To this end, we reconstructed the consensus of all the sequences used in this analysis, namely the two $K^+$ and three $Na^+$ and $Ca^{2+}$ channels.

The total of eight sequences (the original sequences and the three consensus sequences) comprised our dataset for probing the evolutionary origins of the voltage-gated ion channel superfamily. To verify the authenticity of these sequences as representatives of their respective families, we examined the effectiveness of the consensus sequences as BLAST queries. This analysis is detailed below.

1. The $Ca^{2+}/Na^+$ channel consensus returned exclusively $Ca^{2+}$ and $Na^+$ channels. The individual sequences making up the consensus sequence were retrieved first followed by many eukaryotic $Ca^{2+}$ and $Na^+$ channels. The first eukaryotic channel was a mouse $Ca^{2+}$ channel. The first false positive $K^+$ channel had a rank of 800.

2. The $K^+$ channel consensus sequence returned many prokaryotic $K^+$ channel sequences, including the sequences making up the consensus sequence. The first Eukarya channel is a human Shaker channel, with a rank of 1000.

3. The overall consensus sequence, i.e. the consensus of $Na^+$, $Ca^{2+}$ and $K^+$ channels, elicited the calcium and sodium channels of this analysis, followed by those of magnetococcus and microbulbifer, then by the $K^+$ channels of this analysis. After a few more prokaryotic $K^+$ channels, the true positive of Eukarya is seen with a rank of 22. It is a Na channel from fruit fly. After this, the true positives are almost all eukaryotic $Na^+$, $Ca^{2+}$ or $K^+$ channel. This particular analysis suggested that searching the sequence database for more prokaryotic $Ca^{2+}$ and $Na^+$ channels was unlikely to have turned up many new sequences.

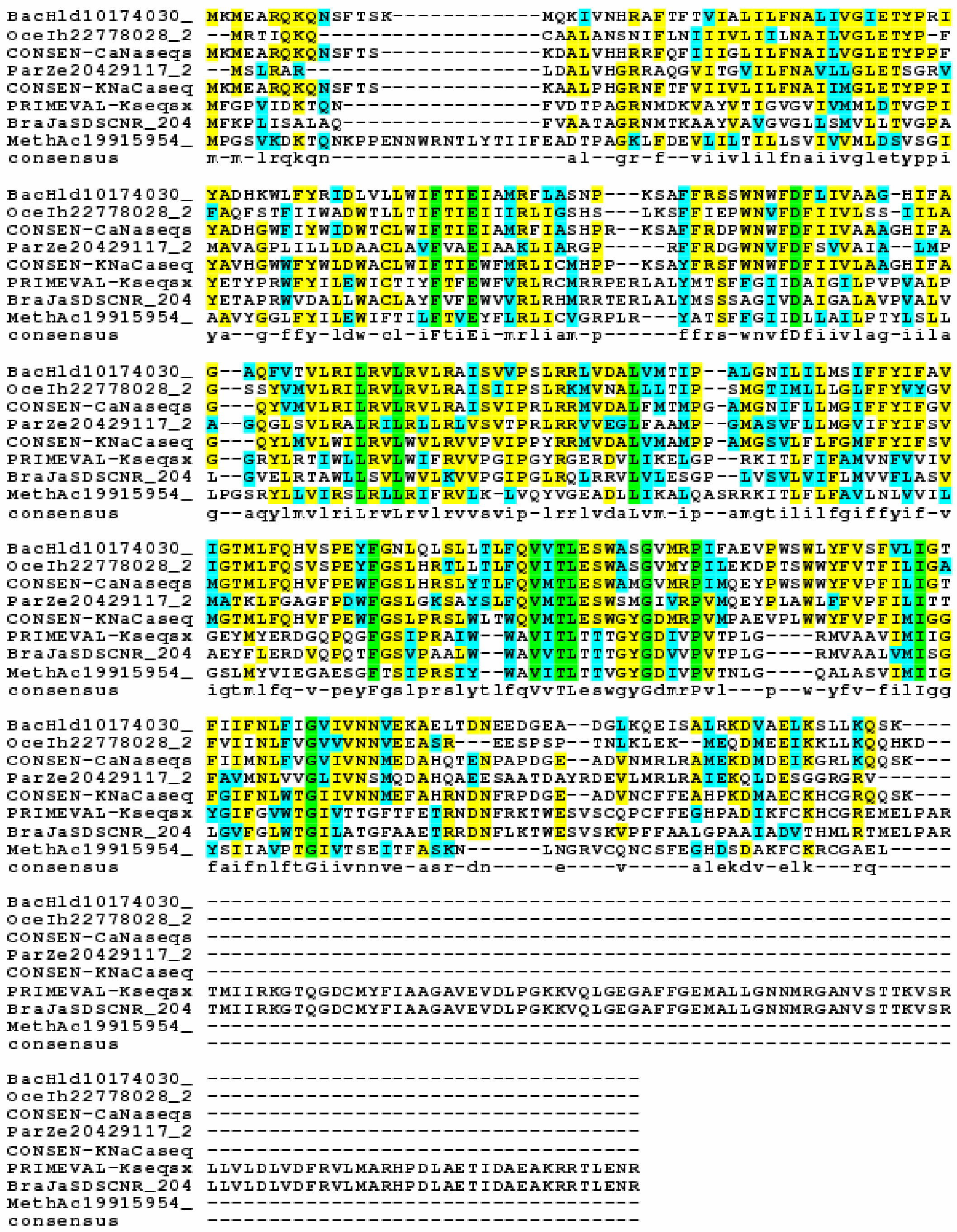

Figure 9. Multiple sequence alignment of the five prokaryotic voltage-gated ion channel sequences and the three consensus sequences that were derived from this group. Note the conservation in the voltage sensor. Around the $K^+$ channel selectivity filter, the first Thr is conserved in all sequences. Similarly the second Gly of the selectivity filter is conserved. Note the conservation of the Gly hinge in all the ion channels. Note the Glutamate in the non-$K^+$ channel sequences, upwards of the $K^+$ channel selectivity filter by a couple of residues.

A multiple sequence alignment was constructed of this dataset using ClustalW, and is shown in figure 9. Using this representative multisequence alignment, we reconstructed the ion channel phylogeny. We computed distances with correction for multiple substitutions and used the NJ algorithm for creating the dendrogram shown in figure 10. 1000 bootstrap trials were performed, and the replicability of each node is also shown. The tree was rooted with the $K^+$ channel consensus sequence outgroup as the primeval channel.

The inclusion of the consensus sequences in the tree enabled the interpretation of the major events in the evolution of the voltage-gated ion channel superfamily. We see that the $Na^+$ channels are the most recent class of the voltage-gated ion channel superfamily. This is evidenced by the monophyletic grouping of the two $Na^+$ channels (67% bootstrap support). The calcium channel of Paracoccus had diverged from an earlier ancestor (79% percent bootstrap support). The $K^+$ channels are seen to represent the earliest class of voltage-gated ion channels. They cluster independently of the sodium and calcium channels, and there is 100% bootstrap support for this observation. A Fitch-Margoliash distance matrix-based tree was also reconstructed to infer the phylogeny of the eight sequences. An identical tree topology was obtained. The uniform preference of this tree topology strengthened the confidence of the predictions.

Note the exceptional conservation of the Gly hinge in all the ion channels. Given the standard deviation in the length of a TM helix is low, and the lengths of the permeation pathway TM helices we have encountered are not unusual, this suggests that there may be features of the gating mechanism that is preserved in the entire voltage-gated ion channel superfamily. A Glutamate residue in the non-$K^+$ channel sequences, upwards of the $K^+$ channel selectivity filter by a couple of residues, suggested that the selection in sodium and calcium channels (whose three-dimensional structure has not been determined yet) might take place nearer to the extracellular mouth of the channel.

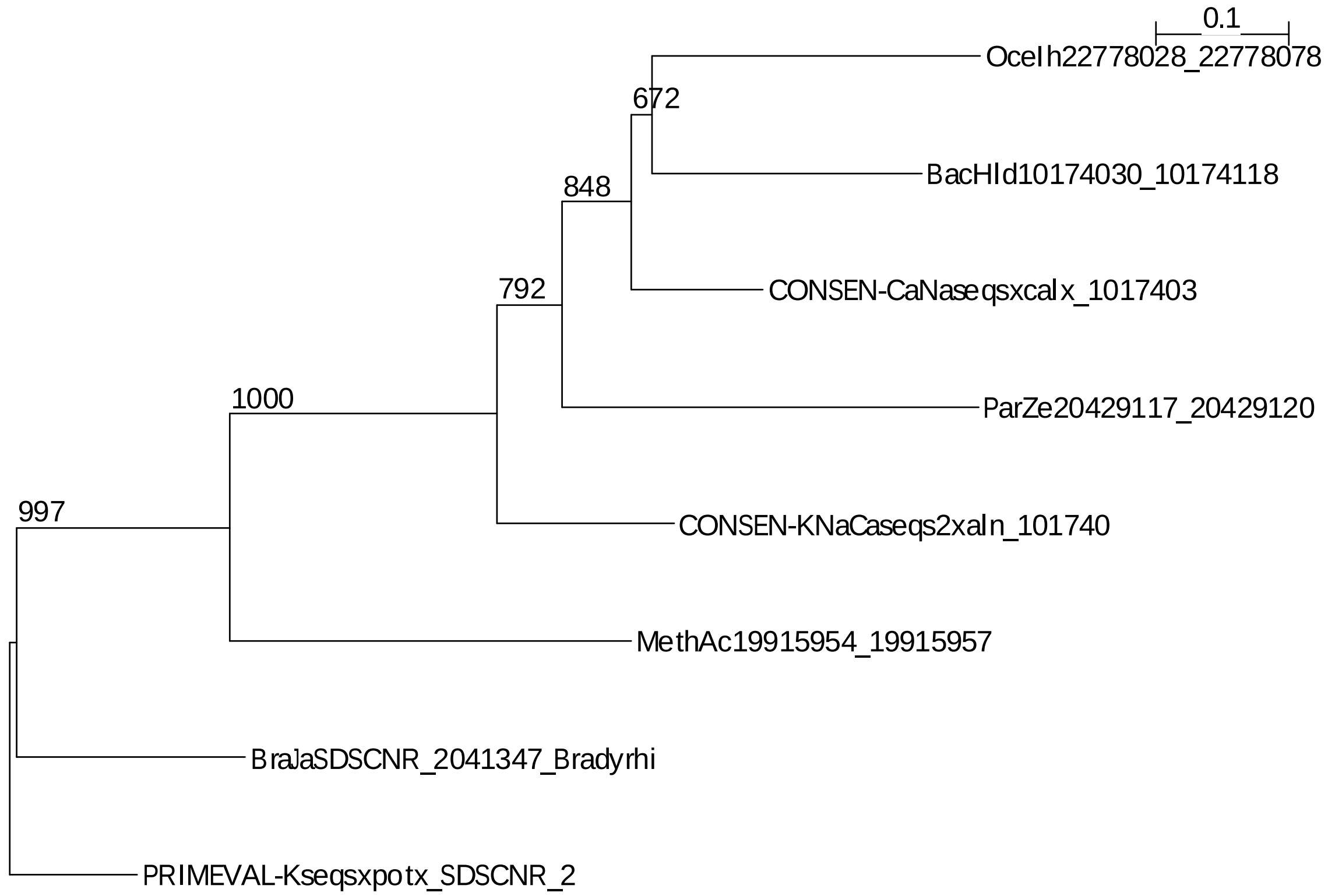

Figure 10. The neighbor-joining tree topology of five prokaryotic voltage-gated ion channels, and three derived consensus sequences. Bootstrap values are given for 1000 trials.

## CONCLUSION

Our discussions provide a template for studying a protein family. $K^+$-channels are a family of membrane ion channel proteins maintaining high heterogeneity in regulation, function, and phyletic subfamily distribution. We observed the expression of a distinct complement of $K^+$-channels in the mouse and the nematode (Palaniappan, 2005).The nematode has a strikingly expanded complement of two-pore channels. Our analysis suggested that the hyperpolarization-activated cyclic-nucleotide-gated potassium channel subfamily was an evolutionary development of later metazoans. Each organism tunes the complement by selective emphasis, and if necessary, contraction, as responses to natural pressure.

*A contribution to linguistics:*

The essentially human faculty of language is typically regarded as “modular”. It is (itself) modeled in a compartmentalised manner, consisting of a central module combined with several other modules. Each module is hypothesized to be inaccessible to other modules, and dedicated to a special function such as feature recognition, moral judgement, and formation of a theory of minds. In a contention analogous to that employed for arguing the physical reality of co-evolution between catalysis and regulation, we suggest that modules must be sufficiently co-evolved to evidence error-free transfer of substantial information. The nature of this co-evolution is realized in physico-chemical properties to provide for the retention of original functions. This theory preserves differentiation and would not negate the manifestation of double dissociation. Significantly, it sets limits to the independence and domain-specificity of modules. The principle of co-evolution might emerge as a fundamental principle with even wider and deeper applications.

## ACKNOWLEDGEMENTS

I would like to thank Dr. C. Grosman (University of Illinois) and Dr. N. Chidambaram (Annamalai University) for valuable discussions. It is a pleasure to thank Prof. Colin Wraight.

```
gi_13540549       -----------seygaavLflLMctfaLiahwlAcIWyaIgnv--------------------erp---ylehkiG-Wlds
gi_15150797       -----------seygaavLMlsMcIfaLnahwlAcIWyaIgnv--------------------erp---yltdkiG-Wlds
gi_3452413        -----------seygaavLflLMctfaLiahwlAcIWyaIgnm--------------------eqp---hmdsriG-Wlhn
gi_4557729        -----------seygaavLflLMctfaLiahwlAcIWyaIgnm--------------------eqp---hmdsriG-Wlhn
gi_12804923       -----------seygaavLflLMctfaLiahwlAcIWyaIgnm--------------------eqp---hmdsriG-Wlhn
gi_11933152       -----------sxygaavLflLMctfaLiahwlAcIWyaIgnm--------------------eqp---hmdsriG-Wlhn
gi_16322995       -----------sqhstivLtlLMsMfaLlahwmAcIWyvIgkm--------------------erednsllkwevG-Wlhe
gi_6912446        -----------sqcsavvLtlLMsVfaLlahwmAcIWyvIgrr--------------------emeandpllwdiG-Wlhe
gi_6331348        -----------sqysavvLtlLMaVfaLlahwvAcVWfyIgqr--------------------eiesseselpeiG-Wlqe
gi_3790565        -----------IeygaavLVlLVcVfgLaahwmAcIWysIgdy--------------------eifdedtktirnnsWlyq
gi_4504831        -----------IeygaavLVlLVcVfgLaahwmAcIWysIgdy--------------------eifdedtktirnnsWlyq
gi_20799308       -----------LeygaavLVlLVcVfgLvahwlAcIWysIgdy--------------------evidevtntiqidsWlyq
gi_4502915        ----------tnypnIfrIsnLVMyIVIiIhwnAcVFysIsk---------------------------aigfgndtWvyp
gi_346350         ----------tnypnIfrIsnLVMyIVIiIhwnAcVFysIsk---------------------------aigfgndtWvyp
gi_7441594        ----------tnypnMfrIgnLVLyILIiIhwnAcIYfaIsk---------------------------figfgtdsWvyp
gi_17485744       ----------tnypnIfrIsnLVLyILViIhwnAcIYyaIsk---------------------------sigfgvdtWvyp
gi_20561127       ----------taypnafrIakLMLyIfVvIhwnscLYfaIsr---------------------------ylgfgrdaWvyp
gi_1518639        ----------lskayVyrVIrttayLLysLhLnscLYywasa---------------------------yqglgsthWvyd
gi_9247066        ----------mdkayIyrVIrttgyLLfiLhInAcVYywasn---------------------------yegigttrWvyd
gi_4885407        ----ifhmtydLasaVvrIVnLIgMMLLlchwdGcLqflVpm---------------------------lqdfpddcWvSi
gi_7959337        ----ifhmtydLasaVvrIfnLIgMMLLlchwdGcLqflVpm---------------------------lqdfppdcWvSi
gi_3168874        ----ifhmtydLasaVvrIfnLIgMMLLlchwdGcLqflVpl---------------------------lqdfppdcWvSl
gi_14762741       ----ifhmtydLasaVmrIcnLIsMMLLlchwdGcLqflVpm---------------------------lqdfprncWvSi
gi_20483928       ----ifhmtydLasaVmrfcnLIsMMLLlchwdGcLqflVam---------------------------lqdfpcncWvSi
gi_487428         -------------------------------nsGdpWenf--------------------------------qnnqalT--
gi_20544687       --tsnsiklvnLlsifistwltaagfIhlVensGdpWenf--------------------------------qnnqalT--
gi_20381097       --tsnsvkfskLlsiIlstwftaagfIhlVensGdpWlkg--------------------------------rnsqniS--
gi_17453679       --tsnsvkfskLlsiIlstwftaagfIhlVensGdpWlkg--------------------------------rnsqniS--
gi_7243225        ilrtqsamfnqVl-iLfctLlcLV-ftg---tcGIqhle---------------------------------ragEnlS--
gi_17384429       ilrtqsamfnqVl-iLfctLlcLV-ftggcrtcGIqhle---------------------------------ragEnlS--
gi_20474705       ----qsamfnqVl-iListLlcLI-ftc---IcGIqhle---------------------------------rigkkln--
2/2-gi_16118235   --------------lLqaValgLLVassFVlLpALVlwg----------------------------------lqGDcS--
2/2-gi_16118231   --------------lLqaValgLLVassFVlLpALVlwg----------------------------------lqGDcS--
2/2-gi_16118238   --------------lLqaValgLLVassFVlLpALVlwg----------------------------------lqGDcS--
2/2-gi_16159718   --------------iVhaVLlgfVtVscFffIpAaVFSv----------------------------------ledDWn--
2/2-gi_13124108   --------------cwhlValLgVVVtVcflVpAVIFah----------------------------------leEaWS--
2/2-gi_20143946   --------------vIstILfILagcIVFVtIpAVIFky----------------------------------ieG-WT--
2/2-gi_10863961   --------------vIstILfILagcIVFVtIpAVIFky----------------------------------ieG-WT--
2/2-gi_20143944   --------------vIstILfILagcIVFVtIpAVIFky----------------------------------ieG-WT--
2/2-gi_13124054   --------------iIstIIfILfgcVLFVaLpAIIFkh----------------------------------ieG-WS--
2/2-gi_5712621    --------------iIstIIfILfgcVLFVaLpAIIFkh----------------------------------ieG-WS--
2/2-gi_13925518   --------------vLsaMLfLLIgcLLFVltptfVFcy----------------------------------med-WS--
gi_20481187       ------------------MLfLLIgcLLFVltptfVFcy----------------------------------med-WS--
2/2-gi_14149764   --------------vLglaLfLtLgtLViLifppMVFSh----------------------------------veG-WS--
2/2-gi_9988112    --------------vLglaLfLtLgtLViLifppMVFSh----------------------------------veG-WS--
2/2-gi_19343981   -------------wlagsgalLs-gLLLFLlLppLLFSh----------------------------------meG-WS--
2/2-gi_13507377   -------------wlagsgalLs-gLLLFLlLppLLFSh----------------------------------meG-WS--
2/2-gi_13124055   --------------itctVIfIVwgVLVhLvIppfVFmv----------------------------------teG-Wn--
2/2-gi_4504849    --------------manmVLigffscIstLcIGAaaFSh----------------------------------yeh-WT--
2/2-gi_13431426   --------------menmVtvgffscMgtLcIGAaaFSq----------------------------------ceE-WS--
2/2-gi_15862337   --------------tenlVVagLLacaatLaLGAVaFSh----------------------------------feG-WT--
2/2-gi_14271510   --------------dIplpIiaLIVfayisca-AaIlpf----------------------------------wetqld--
2/2-gi_20549125   --------------dIplpIiaLIVfayisca-AaIlpf----------------------------------wetqld--
2/2-gi_11545761   -------------------IlgLfaVLLscca-saMYTs----------------------------------veG-Wd--
2/2-gi_16306555   -------------------IlctasILIscca-saMYTp----------------------------------ieG-WS--
gi_107366         -----------LghtLrasMreLgLLIfFLfIGvILFSsaVY----------------fAEade---------ptthFqS-
gi_116430         -----------LghtLrasMreLgLLIfFLfIGvILFSsaVY----------------fAEade---------ptthFqS-
gi_13648556       -----------LgktLqasMreLgLLIfFLfIGvILFSsaVY----------------fAEadd---------ddslFpS-
gi_189673         -----------LgqsLkasMreLgLLIfFLfIGvILFSsaVY----------------fAEadd---------ptsgFSS-
gi_18549646       -----------LgqtLkasMreLgLLIfFLfIGvILFSsaVY----------------fAEadd---------ptsgFSS-
gi_1345813        -----------LgqtLkasMreLgLLIfFLfIGvILFSsaVY----------------fAEade---------resqFpS-
gi_11525742       -----------LgqtLkasMreLgLLIfFLfIGvILFSsaVY----------------fAEaee---------aeshFSS-
gi_5031819        -----------LgqtLkasMreLgLLIfFLfIGvILFSsaVY----------------fAEvde---------peshFSS-
gi_14485555       -----------LgqtLrasMreLgLLIfFLfIGvVLFSsaVY----------------fAEvdr---------vdshFTS-
gi_386100         -----------------------------------------------------------AEadn---------qgthFSS-
gi_13648551       -----------LgktLqasMreLgLLIfFLfIGvILFSsaVY----------------fAEadn---------qgthFSS-
gi_13242172       -----------LgftLrrsyneLgLLILFLaMGiMIFSsLVF----------------fAEKde---------datkFTS-
gi_3913257        -----------LgftLrrsyneLgLLILFLaMGiMIFSsLVF----------------fAEKde---------datkFTS-
gi_7671615        -----------LgftLrrsyneLgLLILFLaMGiMIFSsLVF----------------fAEKde---------ddtkFkS-
gi_12313899       -----------LghtLrastnefLLLIIFLaLGvLIFatMIY----------------yAERigarpsdprgndhtDFkn-
gi_18549632       ---------------------------------------MIY----------------yAERigarpsdprgndhtDFkn-
gi_3023493        -----------LghtLrastnefLLLIIFLaLGvLIFatMIY----------------yAERigarpsdprgndhtDFkn-
gi_8488974        -----------LghtLrastnefLLLIIFLaLGvLIFatMIY----------------yAERigadpddilgsnhtyFkn-
gi_21217561       -----------LghtLrastnefLLLIIFLaLGvLIFatMIY----------------yAERvgaqpndpsasehtqFkn-
gi_21217563       -----------LghtLrastnefLLLIIFLaLGvLIFatMIY----------------yAERvgaqpndpsasehtqFkn-
gi_1352085        -----------LghtLrastnefLLLIIFLaLGvLIFatMIY----------------yAERigaqpndpsasehthFkn-
gi_6007795        -----------LgytLkscaseLgfLLfsLtMAiIIFatVMF----------------yAEKgs---------saskFTS-
gi_4826794        -----------LgytLkscaseLgfLLfsLtMAiIIFatVMF----------------yAEKgs---------saskFTS-
gi_9789987        -----------LgytLkscaseLgfLLfsLtMAiIIFatVMF----------------yAEKgs---------saskFTS-
gi_21361266       -----------LgytLkscaseLgfLLfsLtMAiIIFatVMF----------------yAEKgt---------nktnFTn-
gi_20070166       -----------LtyaLkrsfkeLgLLLMYLaVGifVFSaLgY----------------tmEqsh---------petlFkS-
gi_6329973        -----------LgatLkysykeVgLLLLYLsVGisIFSvVaY----------------tiEKee---------neG-laT-
gi_11420867       -----------LgatLkhsyreVgILLLYLaVGvsVFSgVaY----------------tAEKee---------dvG-FnT-
gi_11428466       -----------LgatLrhsyheVgLLLLFLsVGisIFSvLIY----------------svEKdd---------htsslTS-
gi_20381121       -----------LgmtItqcyeeVgLLLLFLsVGisIFStVeY----------------fAEqsi---------pdttFTS-
gi_19070539       -----------fgftLrqcyqqVgcLLLFIaMGiftFSaaVY----------------svEhdv---------pstnFTT-
gi_19070541       -----------LgltVrrctrefgLLLLFLaVAitLFSpLVY----------------vAEKes--------grvlEFTS-
gi_5679601        -----------LgltarrctrefgLLLLFLcVAiaLFapLLY----------------viEnem--------adspEFTS-
gi_6912444        -----------LgltMrrcarefgLLLLFLcVAmaLFapLVh----------------lAERel--------garrDFSS-
gi_19070537       -----------LgltLkrcyreMVMLLVFIcVAmaIFSaLsq----------------llEhgldle-----tsnkDFTS-
gi_6166006        ---------------VyahskeLItawyigfLvlIfaSfLVY----------------lAEKd----------ansDFSS-
```

Figure S1. (continued)

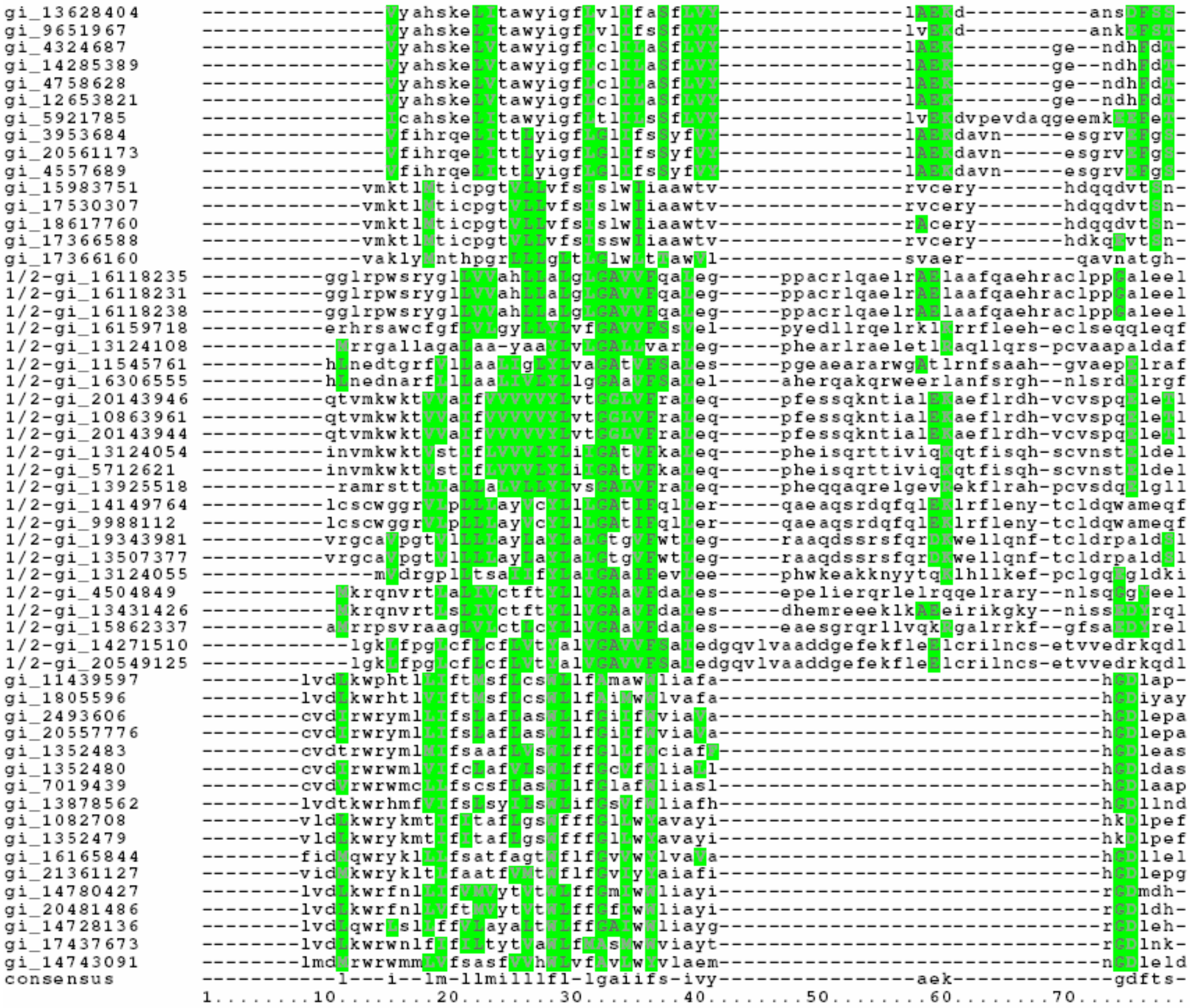

Figure S1. (continued)

```
gi_13540549         lgvqlgkryngsdp---------------------------asgpsvqdkYVtAlYFtfsSLTSVGFGnVsP---------
gi_15150797         lgqqigkryndsds---------------------------ssgpsikdkYVtAlYFtfsSLTSVGFGnVsP---------
gi_3452413          lgdqigkpynssg----------------------------lggpsikdkYVtAlYFtfsSLTSVGFGnVsP---------
gi_4557729          lgdqigkpynssg----------------------------lggpsikdkYVtAlYFtfsSLTSVGFGnVsP---------
gi_12804923         lgdqigkpynssg----------------------------lggpsikdkYVtAlYFtfsSLTSVGFGnVsP---------
gi_11933152         lgdqigkpynssg----------------------------lggpsikdkYVtAlYFtfsSLTSVGFGnVsP---------
gi_16322995         lgkrlespyygnnt---------------------------lggpsirsaYIaAlYFtLsSLTSVGFGnVsa---------
gi_6912446          lgkrlevpyvng-s---------------------------vggpsrrsaYIaAlYFtLsSLTSVGFGnVca---------
gi_6331348          larrletpyylvgrrpaggnssgqsdncsssseangtglellggpslrsaYItslYFALsSLTSVGFGnVsa---------
gi_3790565          lamdigtpyqfngsgsgk----------------------weggpsknsvYIsslYFtMtSLTSVGFGnIaP---------
gi_4504831          lamdigtpyqfngsgsgk----------------------weggpsknsvYIsslYFtMtSLTSVGFGnIaP---------
gi_20799308         lalsigtpyryn-tsagi----------------------weggpskdslYVsslYFtMtSLTTIGFGnIaP---------
gi_4502915          dindpef-------------------------------------grlarkYVyslYWstLTLTTIG-EtppP---------
gi_346350           dindpef-------------------------------------grlarkYVyslYWstLTLTTIG-EtppP---------
gi_7441594          nisipeh-------------------------------------grlsrkYIyslYWstLTLTTIG-EtppP---------
gi_17485744         nitdpey-------------------------------------gylareYIyclYWstLTLTTIG-EtppP---------
gi_20561127         dpaqpgf-------------------------------------erlrrqYLysFYFstLiLTTVG-dtppP---------
gi_1518639          gvgn---------------------------------------------sYIrcYYFAVkTLiTIG-GlpdP---------
gi_9247066          gegn---------------------------------------------eYLrcYYWAVrTLiTIG-GlpeP---------
gi_4885407          nnmvn---------------------------------------nswgkqYsyAlFkAMshMlCIGYGrqaP---------
gi_7959337          nhmvn---------------------------------------hswgrqYshAlFkAMshMlCIGYGqqaP---------
gi_3168874          nemvn---------------------------------------dswgkqYsyAlFkAMshMlCIGYGaqaP---------
gi_14762741         ngmvn---------------------------------------hswselYsfAlFkAMshMlCIGYGrqaP---------
gi_20483928         ngmvn---------------------------------------hwwselYsfAlFkAMshMlCIGYGrqaP---------
gi_487428           --------------------------------------------------YwEcvYllMVTMSTVGYGDVya---------
gi_20544687         --------------------------------------------------YwEcvYllMVTMSTVGYGDVya---------
gi_20381097         --------------------------------------------------YwEsiYlvMaTtSTVGFGDVVa---------
gi_17453679         --------------------------------------------------YfEsiYlvMaTtSTVGFGDVVa---------
gi_7243225          --------------------------------------------------lLtsFYFcIVTfSTVGYGDVtP---------
gi_17384429         --------------------------------------------------lLtsFYFcIVTfSTVGYGDVtP---------
gi_20474705         --------------------------------------------------lfDslYFcIVTfSTVGFGDVtP---------
2/2-gi_16118235     --------------------------------------------------lLgAvYFcfsSLSTIGlEDLLPgrg------
2/2-gi_16118231     --------------------------------------------------lLgAvYFcfsSLSTIGlEDLLPgrg------
2/2-gi_16118238     --------------------------------------------------lLgAvYFcfsSLSTIGlEDLLPgrg------
2/2-gi_16159718     --------------------------------------------------FLEsFYFcfISLSTIGlGDyVPgegy-nqk
2/2-gi_13124108     --------------------------------------------------FLDAFYFcfISLSTIGlGDyVPgeap-gqp
2/2-gi_20143946     --------------------------------------------------aLEsiYFvVVTLTTVGFGDfVaggna-gin
2/2-gi_10863961     --------------------------------------------------aLEsiYFvVVTLTTVGFGDfVaggna-gin
2/2-gi_20143944     --------------------------------------------------aLEsiYFvVVTLTTVGFGDfVaggna-gin
2/2-gi_13124054     --------------------------------------------------aLDAiYFvVITLTTIGFGDyVagg-s-die
2/2-gi_5712621      --------------------------------------------------aLDAiYFvVITLTTIGFGDyVagg-s-die
2/2-gi_13925518     --------------------------------------------------kLEAiYFvIVTLTTVGFGDyVagadp-rqd
gi_20481187         --------------------------------------------------kLEAiYFvIVTLTTVGFGDyVagklg--eg
2/2-gi_14149764     --------------------------------------------------FsEGFYFAfITLSTIGFGDyVvgtdp-skh
2/2-gi_9988112      --------------------------------------------------FsEGFYFAfITLSTIGFGDyVvgtdp-skh
2/2-gi_19343981     --------------------------------------------------YtEGFYFAfITLSTVGFGDyVigmnp-sqr
2/2-gi_13507377     --------------------------------------------------YtEGFYFAfITLSTVGFGDyVigmnp-sqr
2/2-gi_13124055     --------------------------------------------------YIEGlYYsfITISTIGFGDfVagvnp-san
2/2-gi_4504849      --------------------------------------------------FfqAYYYcfITLTTIGFGDyValqkdqalq
2/2-gi_13431426     --------------------------------------------------FfhAYYYcfITLTTIGFGDyValqtkgalq
2/2-gi_15862337     --------------------------------------------------FfhAYYYcfITLTTIGFGDfValqsgealq
2/2-gi_14271510     --------------------------------------------------FenAFYFcfVTLTTIGFGD---------tvl
2/2-gi_20549125     --------------------------------------------------FenAFYFcfVTLTTIGFGD---------tvl
2/2-gi_11545761     --------------------------------------------------YVDslYFcfVTfSTIGFGDLVs-sqhaayr
2/2-gi_16306555     --------------------------------------------------YfDslYFcfVafSTIGFGDLVs-sqnahye
gi_107366           --------------------------------------------------ipDAFWWAVVTMTTVGYGDMkP---------
gi_116430           --------------------------------------------------ipDAFWWAVVTMTTVGYGDMkP---------
gi_13648556         --------------------------------------------------ipDAFWWAVVTMTTVGYGDMyP---------
gi_189673           --------------------------------------------------ipDAFWWAVVTMTTVGYGDMhP---------
gi_18549646         --------------------------------------------------ipDAFWWAVVTMTTVGYGDMhP---------
gi_1345813          --------------------------------------------------ipDAFWWAVVSMTTVGYGDMVP---------
gi_11525742         --------------------------------------------------ipDAFWWAVVSMTTVGYGDMyP---------
gi_5031819          --------------------------------------------------ipDGFWWAVVTMTTVGYGDMcP---------
gi_14485555         --------------------------------------------------ipEsFWWAVVTMTTVGYGDMaP---------
gi_386100           --------------------------------------------------ipDAFWWAVVTMTTVGYGDMrP---------
gi_13648551         --------------------------------------------------ipDAFWWAVVTMTTVGYGDMrP---------
gi_13242172         --------------------------------------------------ipasFWWAtITMTTVGYGDIyP---------
gi_3913257          --------------------------------------------------ipasFWWAtITMTTVGYGDIyP---------
gi_7671615          --------------------------------------------------ipasFWWAtITMTTVGYGDIyP---------
gi_12313899         --------------------------------------------------ipiGFWWAVVTMTTLGYGDMyP---------
gi_18549632         --------------------------------------------------ipiGFWWAVVTMTTLGYGDMyP---------
gi_3023493          --------------------------------------------------ipiGFWWAVVTMTTLGYGDMyP---------
gi_8488974          --------------------------------------------------ipiGFWWAVVTMTTLGYGDMyP---------
gi_21217561         --------------------------------------------------ipiGFWWAVVTMTTLGYGDMyP---------
gi_21217563         --------------------------------------------------ipiGFWWAVVTMTTLGYGDMyP---------
gi_1352085          --------------------------------------------------ipiGFWWAVVTMTTLGYGDMyP---------
gi_6007795          --------------------------------------------------ipasFWYtIVTMTTLGYGDMVP---------
gi_4826794          --------------------------------------------------ipasFWYtIVTMTTLGYGDMVl---------
gi_9789987          --------------------------------------------------ipaAFWYtIVTMTTLGYGDMVP---------
gi_21361266         --------------------------------------------------ipaAFWYtIVTMTTLGYGDMVP---------
gi_20070166         --------------------------------------------------ipqsFWWAIITMTTVGYGDIyP---------
gi_6329973          --------------------------------------------------ipacWWWAtVSMTTVGYGDVVP---------
gi_11420867         --------------------------------------------------ipacWWWGtVSMTTVGYGDVVP---------
gi_11428466         --------------------------------------------------ipicWWWAtISMTTVGYGDthP---------
gi_20381121         --------------------------------------------------ipcAWWWAttSMTTVGYGDIrP---------
gi_19070539         --------------------------------------------------iphsWWWAaVSISTVGYGDMyP---------
gi_19070541         --------------------------------------------------ipasYWWAIISMTTVGYGDMVP---------
gi_5679601          --------------------------------------------------ipacYWWAVITMTTVGYGDMVP---------
gi_6912444          --------------------------------------------------ipasYWWAVISMTTVGYGDMVP---------
gi_19070537         --------------------------------------------------ipaAcWWvIISMTTVGYGDMyP---------
gi_6166006          --------------------------------------------------YaDslWWGtITLTTIGYGDktP---------
```

Figure S1. (continued)

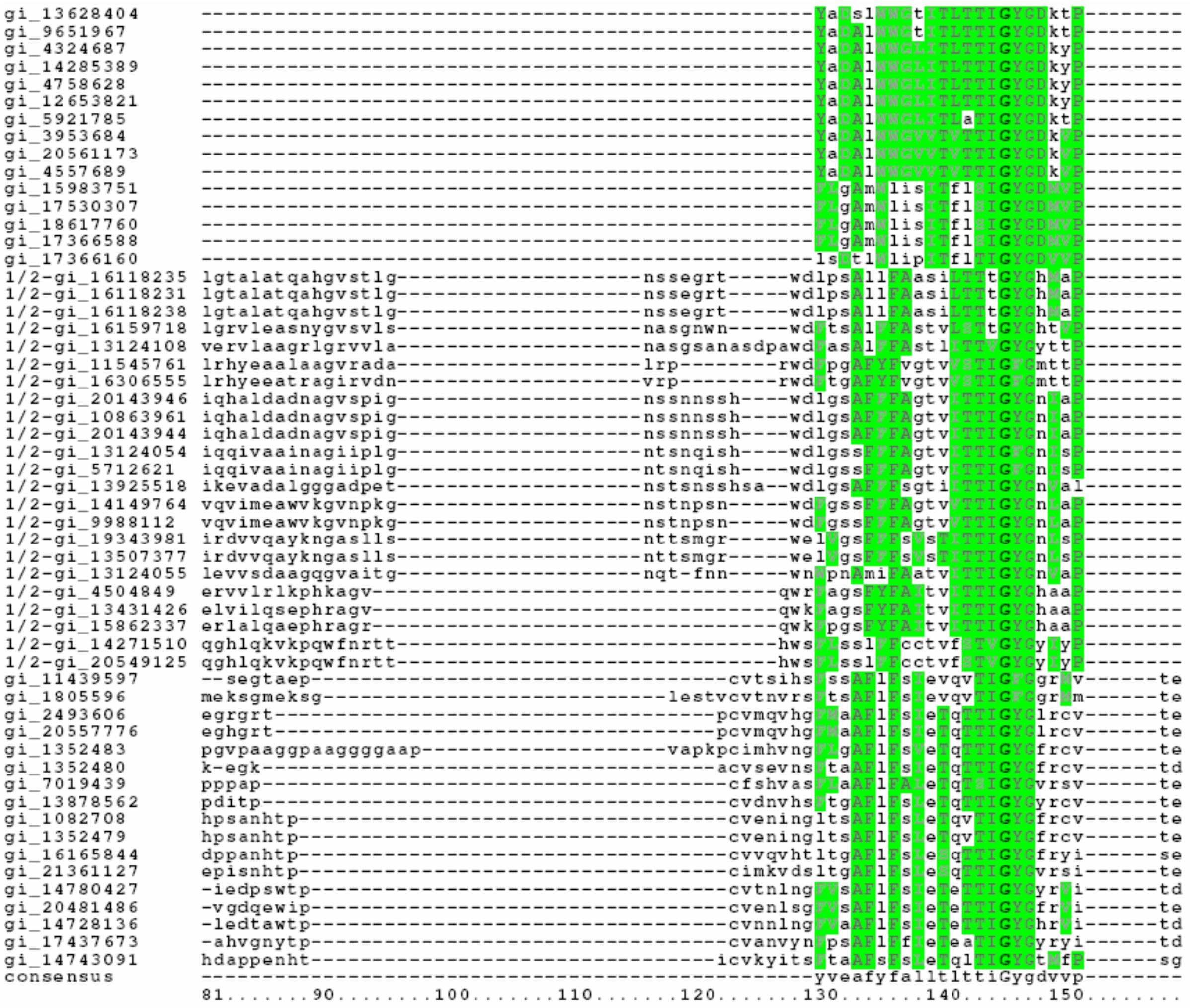

Figure S1. (continued)

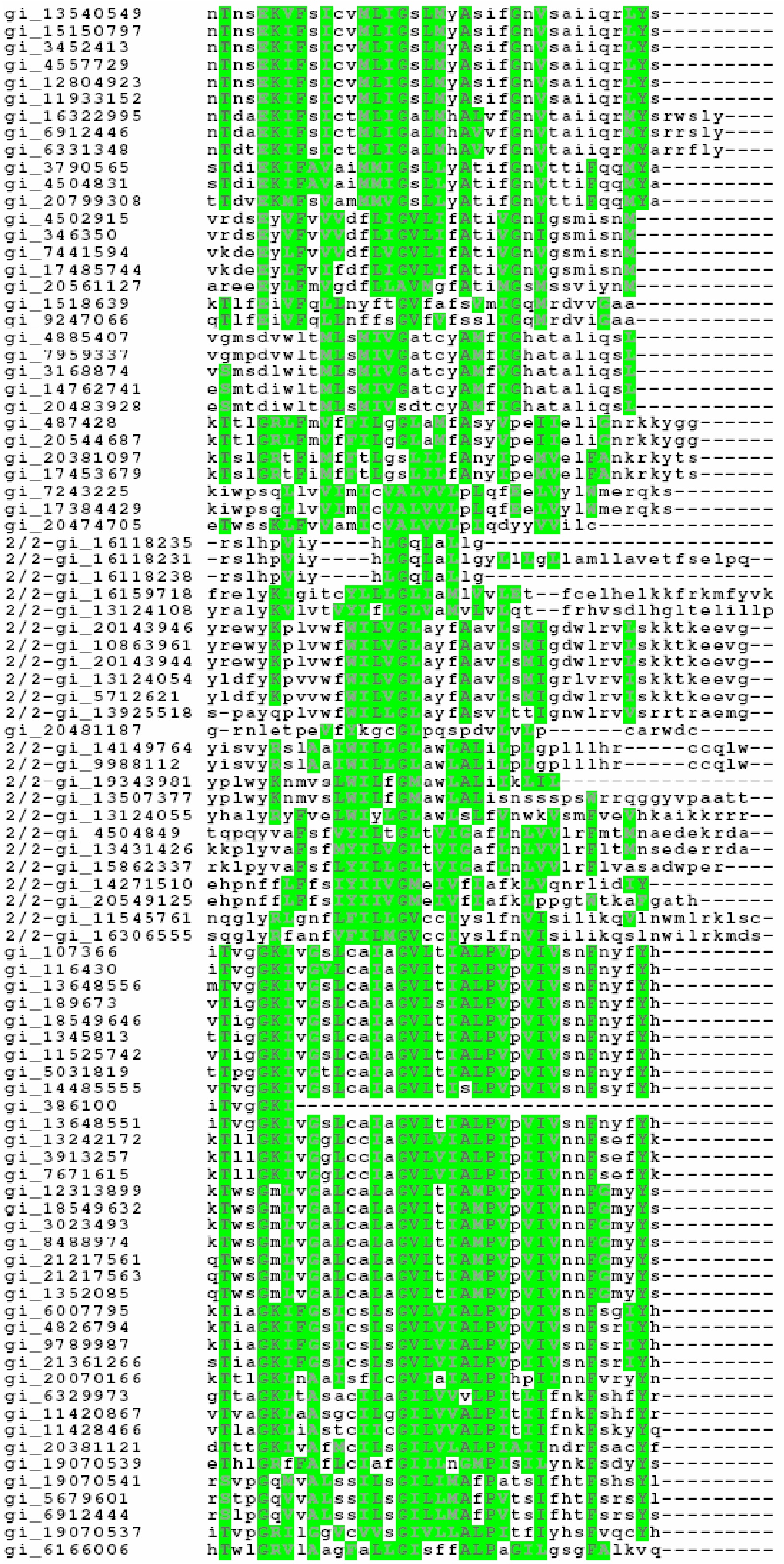

Figure S1. (continued)

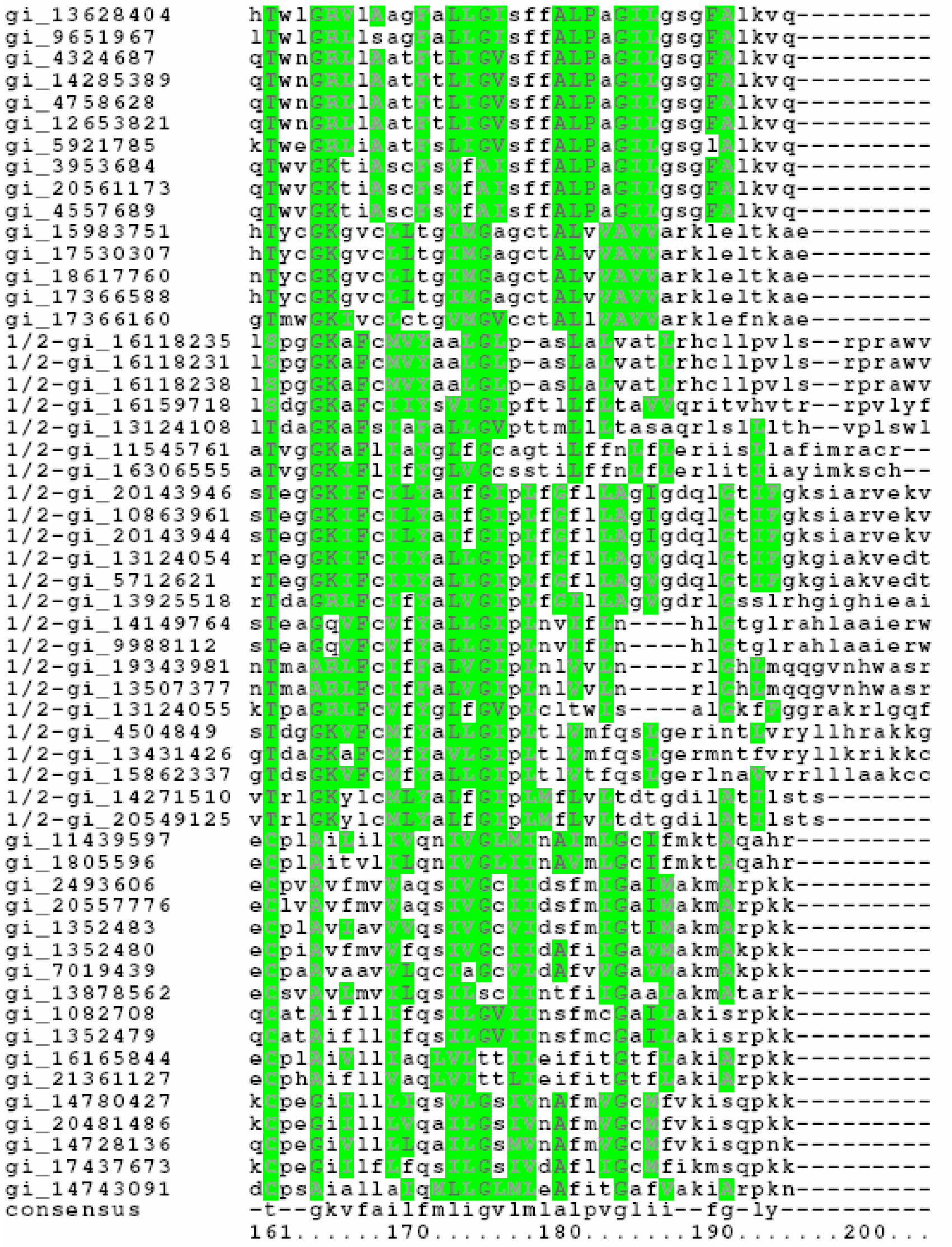

Figure S1. Multiple alignment of 146 permeation pathways of 99% non-redundant K+ channels in human, designated by their Genbank gi numbers. The 23 two-pore channels contribute two permeation pathways each and these are designated as 1/2-gi_number and 2/2-gi_number respectively. The conservation information in the alignment is represented as follows: Green indicates that there is at least 40% similarity for the residue in a particular column. If a residue appears in upper case, then stronger the letter appears, higher is the corresponding residue conservation. A residue in lower case denotes that it is dissimilar to the other residues in the column. The residue consensus is given beneath the alignment segments and is in upper-case if a column is all-identical and lower case if a column is at least 40% similar (which can aid in identifying transmembrane segments). The alignment residue numbering is given below the consensus.